\documentclass[onecolumn,traditabstract]{aa} 

\usepackage{graphicx}
\usepackage{pstricks}
\usepackage{txfonts}
\usepackage{lscape}
\usepackage{longtable,lscape}

\usepackage{natbib}
\bibpunct{(}{)}{;}{a}{}{,} 

\usepackage{hyperref}

\begin{document}
   \title{Colours of minor bodies in the outer solar system}

   \subtitle{II: A statistical analysis revisited}

   \author{
        O.~R.~Hainaut  \inst{1} \and 
        H.~ Boehnhardt \inst{2} \and
        S.~Protopapa   \inst{3}
      }

   \institute{
     European Southern Observatory (ESO), 
     Karl Schwarzschild Stra{\ss}e, 
     85\,748 Garching bei M\"unchen, Germany \\  \email{ohainaut@eso.org}
     \and
      Max-Planck-Institut f\"ur Sonnensystemforschung, 
      Max-Planck Stra\ss e 2, 37\,191 Katlenburg-Lindau, Germany       
      \and
      Department of Astronomy, University of Maryland, 
      College Park, MD 20\,742-2421, USA
}

   \date{Received May 2012; accepted July 2012}

 

\abstract{

We present an update of the visible and near-infrared colour database
of Minor Bodies in the Outer Solar System (MBOSSes), which now
includes over 2000 measurement epochs of 555~objects, extracted from
over 100 articles. The list is fairly complete as of December 2011.
The database is now large enough to enable any dataset with a large
dispersion to be safely identified and rejected from the analysis. The
selection method used is quite insensitive to individual
outliers. Most of the rejected datasets were observed during the early
days of MBOSS photometry. The individual measurements are combined in
a way that avoids  possible rotational artifacts.  The spectral gradient
over the visible range is derived from the colours, as well as the $R$
absolute magnitude $M(1,1)$.  The average colours, absolute magnitude,
and spectral gradient are listed for each object, as well as the
physico-dynamical classes using a classification adapted from Gladman
and collaborators.
Colour-colour diagrams, histograms, and various other plots are
presented to illustrate and investigate class characteristics and
trends with other parameters, whose significances are evaluated using
standard statistical tests.
Except for a small discrepancy for the $J-H$ colour, the largest
objects, with $M(1,1)<5$, are indistinguishable from the smaller
ones. The larger ones are slightly bluer than the smaller ones in $J-H$. 
Short-period comets, Plutinos and other resonant objects, hot
classical disk objects, scattered disk objects and detached disk
objects have similar properties in the visible, while the cold
classical disk objects and the Jupiter Trojans form two separate
groups of their spectral properties in the visible wavelength range.
The well-known colour bimodality of Centaurs is confirmed. 
The hot classical disk objects with large inclinations, or large
orbital excitations are found to be bluer than the others, confirming
a previously known result. Additionally, the hot classical disk
objects with a smaller perihelion distance are bluer than those that
do not come as close to the Sun.
The bluer hot classical disk objects and resonant objects have fainter
absolute magnitudes than the redder ones of the same class.
Finally, we discuss possible scenarios for the origin of the colour
diversity observed in MBOSSes, i.e. colouration caused by evolutionary 
or formation processes.


The colour tables \footnote{ Tables 2 and 3 are only available in
  electronic form at the CDS via anonymous ftp to cdsarc.u-strasbg.fr
  (130.79.128.5) or via
  \href{http://cdsweb.u-strasbg.fr/cgi-bin/qcat?J/A+A/} {\tt
    http://cdsweb.u-strasbg.fr/cgi-bin/qcat?J/A+A/}.  } and all plots
are also available on the MBOSS colour web page\footnote{
  \href{http://www.eso.org/~ohainaut/MBOSS} {\tt
    http://www.eso.org/\~{}ohainaut/MBOSS} }, which will be updated
when new measurements are published

}


   \maketitle
%


\section{Introduction}

Minor bodies in the outer solar system (MBOSSes) comprise objects in
the Kuiper Belt (KB) and more generally in the Transneptunian (TN)
region, as well as small bodies in the giant planets region that came
from either the KB or TN regions, but are now no longer immediate members of
these environments. Centaurs are considered as escapees
\citep{GMV08,Kav+08}, scattered from the KB and TN environments towards
the Sun, possibly also representing a major source of short-period
comets (SPCs). Some MBOSSes may also be found among the satellites of
the giant planets since they might have been stranded there by the
gravitational capturing from the Centaur population \citep{DL97}. 
Another large population of MBOSSes can be found as long-period (LPCs)
or Oort Cloud comets. These objects are believed to have formed in the
region of the giant planets and been scattered afterwards into the
very distant domains of the solar system \citep{Don+04}, although it
cannot presently be excluded that contamination by extra-solar comets
\citep{Lev+10} may exist among the Oort Cloud population.

The MBOSSes found in the planetary system environment formed and have
evolved in the region of the giant planets.  They are considered as
remnants of the planetesimal population, but their current distance
range is not necessarily that of their formation.  Nonetheless,
MBOSSes may contain valuable information about the environment and the
physical conditions at the time of their formation.
On the other hand, their prolonged presence in the region of the
distant planets and beyond may have caused alterations to their
physical properties. For instance, high energy and particle radiation
is suspected to modify the colours and albedos of materials in space
\citep{deB+08} and collisions can affect the body either as a whole
(fragmentation or growth) or in parts (resurfacing with excavated
material). In addition, it cannot be excluded that intrinsic activity may
have changed the surface constitution of the MBOSSes, since a number
of ices, believed or known to be present in these objects, can
sublimate at very large distances from the Sun \citep[for instance
  $CH_4$, CO, and $N_2$ up to about 45, 65, and 80AU,
  respectively;][]{Del82, MS04}.  Last, but not least, larger MBOSSes
may have experienced some alteration to their internal structure
\citep{McK+08}. It is thus of interest to characterize the population
properties and  explore possible connections with the origin of the
bodies and/or their evolutionary pathways in the solar system.

With this study, we focus on the photometric properties of MBOSSes and
the characterization of the different dynamical populations among
them. The photometric properties comprise published measurements of
their photometric brightnesses, filter colours, and spectral slopes of
reflected continuum light in the visible and near-infrared (NIR)
wavelength regions \citep{Dor+08}. The brightness of an object is a
first indicator on its size and albedo; filter colours allow a
coarse characterization of the spectral energy distribution of its
reflected light and may provide constraints on its surface properties,
i.e. its wavelength-dependent reflectivity.

The visible and NIR wavelength region up to 5$\mu$m contains
surface-reflected sunlight, i.e. essentially either a bluish, neutral,
or a reddened solar spectrum, occasionally with imprinted absorption
features from specific surface materials. Continuum colours and
gradients are well defined, perhaps with the exception of the $H K L
M$ bands for objects with very strong ice absorption features
\citep{TSS11}. The photometric parameters of MBOSSes and their
correlations with then dynamical and other properties of the bodies
were analysed in the past using various different data-sets and
statistical methods \citep[for a review, see] [and references
  therein]{Dor+08}.  The key findings were that the TNOs cover a wide
range of colours and spectral slopes in the visible from slightly
bluish ($-5$ to $-10$\%/100~nm) to very red ($40$ to $55\%$/100~nm),
while in the NIR they display a fairly narrow dispersion around solar
colours.  Differences in the wavelength-dependent surface reflectivity
were noted for dynamically hot and cold Classical Disk Objects (CDOs),
with the latter representing a very red population in the solar
system. Among the hot CDOs, an anti-correlation with high significance
between the surface reddening on one hand and inclination and
eccentricity on the other was interpreted as an indicator of
evolutionary changes due to impacts and cosmic radiation
\citep{TB02,PLJ08}. Centaurs may have a bimodal colour distribution
with a neutral to slightly red and a very red sub-population
\citep{Teg+08}. A number of weaker correlations between photometric
and dynamical properties of KBOs are addressed in the
literature. However, no clear convincing picture with quantitative
modelling results has evolved, although various attempts to obtain a
qualitative understanding of how the surface colours were produced as
the result of evolutionary processes at the surface in the outer space
environment have been published \citep[for a brief review,
  see][]{Dor+08}.

{A completely independent approach to MBOSS colour statistics was
  introduced by \citet{Bar+05b}, where the objects were grouped based
  on their $B, V, R, I, J$ colours to form taxonomic classes. The
  colour study was performed using a principal component (PC) analysis
  \citep{RJ93} and with the G-mode multivariate statistics
  \citep{Cor77,Ful+00}. \cite{Ful+08} describe in detail the method
  and its main results. In the PC analysis,  the original $n$ colours
  are replaced by the minimum possible number of linear combinations
  of colours (the $m$ principal components) required to reproduce the
  original distribution, the $n-m$ additional ones containing only
  noise. They found that most of the MBOSS colour range is accounted
  for by a single variable, closely related to the overall slope of
  the reflectivity spectrum over the visible range (which we define
  quantitatively in Sect.~\ref{sect:dbcontent}). The G-mode analysis
  reveals two main classes, labeled BB and RR, containing the
  neutral-grey and very red objects, respectively. The intermediate
  objects sort themselves into two additional classes, BR and
  IR. \cite{Per+10} published an updated list of the taxonomic classes
  for 151 objects, and studied their distributions over the dynamical
  classes and other parameters.  In the 1970s-1980s, the G-mode-based
  taxonomic classification of main belt asteroids was very successful
  \citep{Tho84,TB89,BF90}, the resulting classes becoming the basis
  for the physico-dynamical understanding of these objects, which is
  also the ultimate goal of the ongoing MBOSS taxonomic
  classification.}

In many cases, the published colour and spectral gradient analyses
rely on different data-sets collected at different telescopes and
instruments and different groups. The sample sizes have grown with
time reaching more than 100 TNOs and Centaurs. A summarizing database
of MBOSS colours comprising data from many different papers were first
compiled and analysed by \citet[hereafter Paper~I]{HD02} and was
publicly accessible through the Internet; it is still available from
the MBOSS web site, but is superseded by this work.

In this paper, we provide a version of this MBOSS colour database that
has been updated both in terms of populations and collected data. It
can be used either for population-wide analysis, as done in this
paper, or to support other works on specific objects or group of
objects either relying on the averaged data presented here (e.g. to
get the colour of a given object) or by going back to the original
publications listed for each object. We present new statistical
analyses of the MBOSS populations based on an enhanced database of
photometric brightnesses and colour measurements of the objects
published in the literature and applying qualified selection criteria
to the data. The qualifying criterion of the data are described in
Section~\ref{REF-2}. Section~\ref{REF-3} introduces the statistical
methods and their application goals for the MBOSS analysis and
outlines the results achieved. Possible interpretations of the
findings related to their formation and evolutionary pathways are
discussed in Section~\ref{REF-4}. The paper ends with a brief summary
of both the major findings and open and controversial issues, as well
as prospects for their clarification in Section~\ref{REF-5}.


\section{Description of the database}\label{REF-2}


The database collects photometric information about MBOSSes, namely of
objects in the Kuiper Belt (KBOs) and in the immediate Transneptunian
region, of Centaurs, of short- and long-period comets (SPCs and LPCs),
and of Jupiter Trojans. It does not contain information on satellites
of the giant planets as far as they are considered to be captured
MBOSSes, nor of Trojans of other planets. The objects are listed under
their current official designation, i.e. number and name if available,
or number and provisional designations, or provisional designation
only.

For objects presenting cometary activity, the measurements collected
here refer to the nucleus only, not to the dust and gas component. In the
case of objects with satellites, measurements for the whole system are
listed under the main designation (e.g. 26308 = 1998SM165), while an
individual member of the system is indicated by a suffix (+B for the
first satellite, for instance 134340-Pluto+B refers to Charon). In the
statistical analysis, the satellites are not given a special status:
they just count as individual objects. 

The photometric information in the database comprises magnitudes or
filter colours in the $UBVRIJHKL$ 
broadband filter system for the visible and the
NIR. In rare cases also spectral gradients and/or colours
measured from spectroscopic data of MBOSSes were entered, relying on
the spectrum to colours conversion presented in the original paper.


All “flavours” of the main filters are considered together without any
conversion. For instance, in this database, Bessell $R$, Kron-Cousin
$R$ and $r$' filters are directly listed as $R$. 
Most authors calibrate their system using \cite{Lan92}, thereby
naturally unifying these subtly different systems.  More exotic
filters are included in the database, but not used in this study.


The literature is searched for relevant papers using the SAO/NASA
ADS\footnote{\href{http://adswww.harvard.edu/}{\tt
    http://adswww.harvard.edu}}, the distant EKO
newsletter\footnote{\href{http://www.boulder.swri.edu/ekonews/}{\tt
    http://www.boulder.swri.edu/ekonews}}, the \cite{Nee11}
compilation of TNO and Centaurs at the Planetary Data
System\footnote{\href{http://sbn.psi.edu/pds/resource/tnocencol.html}{\tt
    http://sbn.psi.edu/pds/resource/tnocencol.html}}, and the astro-ph
preprints from
arXiv\footnote{\href{http://arxiv.org/archive/astro-ph}{\tt
    http://arxiv.org/archive/astro-ph}}.

Only magnitude measurements that are explicitly presented as
“simultaneous” are considered as a single epoch. This means that the
individual filters must have been observed in a sequence, so that no
more than $\sim$1h elapsed during a colour measurement. The only exception
was for average colours obtained from full light-curves. By default,
spectral gradients and colours obtained from spectroscopy fulfill the
``simultaneity'' criterion. Photometric measurements not obtained 
simultaneously are listed as separate data, and are not used for the
average colour estimates. The method used to carefully combine
measurements from different epochs without introducing colour
artifacts from a possible lightcurve is the same as in paper~I.  

The original database, described in paper~I, contained information for
only 104 objects. We had neither the means to evaluate the quality of,
nor the luxury to reject some of the measurements. The online version
continuously grew, and the current MBOSS database includes over 550
objects, with over 2000 measurements extracted from over 100
papers. We can now afford to reject some measurements based upon a
careful and quantifiable approach.  The process of cleaning data
should be approached with caution, as it could potentially select
against special or interesting objects. We developed a method to
evaluate the quality of a data-set that is robust against genuine
outliers. As first noted by \cite{HBO+01} and discussed later in the
present paper, most objects have a fairly linear reflectivity spectrum
over the visible wavelength range; we used this to assess the quality
of a set of measurements. For each individual measurement, the
distance between the data point in a colour-colour diagram and the
“reddening line” (locus of the objects with a linear reflection
spectrum, see Section \ref{sect:reddening}) was computed as
follows. We defined $(Cx,Cy)$ to be the colours of the object, and $(Cx, Fy)$,
$(Fx, Cy)$ the two points on the reddening line that have the same $x$
and $y$ as the data point, respectively. The distance estimator was
defined as
\begin{equation}
d = \sqrt{ \sum_{i=1}^{n} \frac{(C_i-F_i)^2}{n}} \ ,
\end{equation}
where the $C_i$ and $F_i$ terms are the coordinates in the $B-V$/$V-R$ and
$V-R$/$R-I$ colour-colour diagrams where available, and $n$ the number of
coordinates available (2 or 4). The overall quality estimator for a
data-set is  
\begin{equation}
D = \sqrt{ \sum_{k=1}^{N}{ \frac{ d_k^2 }{ N}}} \ ,$$ 
\end{equation}
where $d_k$ are the individual distances and $N$ the number of
measurement sets in the considered paper. $D$ behaves like an error in
the photometry, expressed in magnitudes. Each of the papers with
$D>0.25$ were scrutinized --this threshold was selected visually as a
$2 \sigma$-like limit in the plot.  The faintest objects were flagged
out and $D$ re-estimated. If $D$ remained above 0.25, the process was
iterated. In some cases, this brought $D$ below 0.25, and the
remaining measurements were retained, assuming that those rejected
were affected by low signal-to-noise ratio. In other cases, the
dispersion in the measurements was uncorrelated with magnitude and the
full data-set was rejected; we consider that a problem affected that
whole paper. This was the case for some of the early literature on
MBOSS photometry. Before rejecting a data-set or a measurements, the
colours were compared with other measurements of the same object, if
available, so as to preserve objects with intrinsic unusual
spectra. In addition, the majority of the objects 
have linear spectra, so even a few objects with non-linear spectra
did not strongly affect the overall $D$ estimator of a paper (thanks to the
quadratic average), and these objects were then preserved in the
database.

This method obviously worked only for measurements in the visible
range. Among the 107 papers considered, 33 did not have suitable data,
and 7 were globally rejected. For each data-set (including IR-only
papers), outliers were individually considered. Only the points that
we have a strong reason to believe were affected by a problem were
rejected (for instance where a note in the paper reported a
problem). In addition, some papers that concentrated on objects with
exceptional spectra were obviously preserved. We believe that this
statistical approach to the cleaning process with a careful and
conservative {\it a posteriori} assessment ensures that the global
data-set is of higher quality.


The database presented in this paper is available
online\footnote{\href{http://www.eso.org/~ohainaut/MBOSS/MBOSS2.dat}{\tt
    http://www.eso.org/\~{}ohainaut/MBOSS/MBOSS2.dat}}, together with
all the plots related to this static version of the database, as a
reference. In parallel, as the database is continually evolving with
the addition of new measurements, another up-to-date
version\footnote{\href{http://www.eso.org/~ohainaut/MBOSS}{\tt
    http://www.eso.org/\~{}ohainaut/MBOSS}} is available, with all
plots and tables. The former should be used only in direct reference
to this paper, while the dynamically updated version can be used for
further generic studies of MBOSSes. The updated versions are produced
in exactly the same way as the static one presented in this paper, with additional
objects and measurements processed as described above.

\subsection{Content of the database} \label{sect:dbcontent}


Each record internally lists the object designation as in the original
publication (i.e. typically using the temporary designation for early
papers, and  the final number in later papers). These were converted to the
current official designation, i.e. either number and name when available, 
number and provisional designation, or only provisional designation. 


The internal database lists the measurements for each epoch. One
record is constituted by the name of the target, the epoch of the
measurements, the reference to the original paper, and the list of
measurements for that epoch as they appear in the paper, i.e. as
magnitudes, colours, or a combination of both.


The orbital elements were retrieved from the MPCORB file, which is regularly
updated from the Minor Planets Center (MPC)
website\footnote{\href{http://www.minorplanetcenter.net/iau/Ephemerides/}{\tt http://www.minorplanetcenter.net/iau/Ephemerides}}. These
elements were used to compute the position of the object at the epoch
of the observations (if available). The orbit semi-major axis $a$,
perihelion $q$, eccentricity $e$, inclination $i$, orbital excitation
defined as ${\cal E} = \sqrt{ \sin^2 i + e^2} $, and the helio- and
geocentric distances $r$ and $\Delta$ as well as the solar phase
$\alpha$ at the time of the observations were stored with the
individual measurements.


The measurements were averaged as in paper~I: For a given epoch, the
matrix of all possible colours was populated from the magnitudes and
colours available at that epoch -- but without mixing different epochs. The
average colours for an object were then obtained as a weighted average
of the corresponding colours, using $1/\sigma$ as the weight, where
$\sigma$ was either the reported or propagated error in the individual
colour measurements.


Additionally, for each epoch, we converted the $R$ magnitude (either
the one reported, or obtained from another magnitude together with the
corresponding colour index) into an absolute $M(1,1,\alpha)$ magnitude
with the computed $r$ and $\Delta$ using $M(1,1,\alpha) = R - 5 \log
(r \Delta)$.  As in paper~I, we neither made no assumption about the
solar phase function, nor corrected for the solar phase effect. The
values of 
$M(1,1,\alpha)$ obtained for different epochs were then averaged into a
final absolute $R$ magnitude, which is reported as $M(1,1)$ in
Table~\ref{tab:allColours}. The solar phase effects were not
considered.  The observations were obtained at small solar phase
angles because of the distance of the objects, and because most of the
observations were acquired close to opposition. The colours of the
objects, whose study is at the core of this paper, were even less
affected than the absolute magnitude.
We also did not make any assumption about the albedo of the object,
hence did not convert the absolute magnitudes into diameters.

For each object, the slope of the (very low resolution) reflectivity
spectrum obtained from the average colour indexes \citep{JM86} was computed
as in paper~I, via a linear regression over the $B-V-R-I$ range
(assigning a lower weight to the $B$ reflectivity), where most
objects display a linear reflectivity spectrum \citep{HBO+01}. This
gradient, \cal S, is expressed in percent of reddening per 100~nm.

\subsection{Physico-dynamical classes}

Since the publication of paper~I, the understanding of the dynamical
classes in the outer solar system has greatly progressed. Paper~I used
fairly arbitrary definitions of the dynamical classes; moreover, the
fairly small number of objects in each class did not allow for a
fine-grained classification. In this paper, we used the dynamical
classes defined by \cite{GMV08} (the now so-called SSBN08 classification,
from the book in which it is published). The membership of the object
was allocated using a combination of integration of the orbit over time
together with cut-off in the orbital element space.


For this paper, we relied on the membership list published by
\citet{GMV08}, to which about 200 objects had been added from a list
generated by C. Ejeta and H. Boehnhardt (private communication) in the
context of the Herschel Key Program ``TNOs are Cool''
\citep{Mue+09}. This list was completed by the MPC catalogue of
Jupiter
Trojans\footnote{\href{http://www.minorplanetcenter.net/iau/lists/JupiterTrojans.html}{\tt http://www.minorplanetcenter.net/iau/lists/JupiterTrojans.html}}.

For the objects not included in these lists, we determined the
dynamical class using a simplified method: objects with an obvious
designation (LPCs, SPCs) were first
flagged as such, then for the remaining few objects (only 26 objects in the
current database), we used an algorithm based only on the osculating
orbital elements, without integration of the orbit, directly inspired
by the flow-chart Fig. 1 in \cite{GMV08}. Furthermore, we separated the
classical TNOs between dynamically hot classical disk objects (with $i
\ge 5\degr$) and dynamically cold classical disk objects ($i <
5\degr$).  The arguments for this separation of CDOs come from the
analysis of their orbital properties, although the first indications
of different sub-populations among the classical TNOs 
came from a photometric study of TNOs by \cite{TR00} and were further
evaluated in papers by \cite{TB02} and \cite{Dor+02}. The dynamical
aspects are addressed in \cite{MLG08} and \cite{GMV08}. We chose
an inclination of $5\degr$ for the separation of the two
populations. A split around orbital excitation $\cal E$ = 0.12
gave similar results (only 3 objects among the 89 cold classical TNOs
changed class). Clearly, a sharp cut in $i$, $e$, or $\cal E$ space is a
simplification, since it can be expected that both the cold and hot
populations actually partially overlap. Following \cite{GMV08},
we also distinguished TNOs as scattered disk objects (SDOs) or
detached disk objects (DDOs). The sharp cuts in orbital element space
can again only be seen as an approximation, since the actual boundary
between the classes is expected to have a ``complicated, fractal
structure'' \citep{MLG08}.

The majority of `resonant' objects belongs to the class of
Plutinos in 3:2 resonance with Neptune. Objects belonging to other
resonances are summarized as `Res.others' objects in the analysis
below. Finally, some of the SPCs with Centaur-like
orbital elements, as defined in \cite{GMV08} are re-assigned to that
class, following \cite{Jew09} and \cite{Teg+08}, namely
29P/Schwassmann-Wachmann~1, 30P/Oterma, 165P/LINEAR, 166P/2001~T4,
167P/2004~PY42 174P aka 60\,558 Echeclus, C/2001~M10 (NEAT), P/2004~A1
(LONEOS), and P/2005~T3 (Read).

\subsection{The database}

Table~\ref{tab:dbStats} lists the classes considered and the number of
objects in each class. Table~\ref{tab:allColours} provides examples of
individual objects, listing the object identification and its
dynamical class together with the main average colours, the spectral
gradient $\cal S$, and the absolute $R$ magnitude $M(1,1)$. The full
table is available in the online supplement.

\begin{table}
\caption{Object physico-dynamical classes (see text for the
  definitions) and number of objects in each class, and some
  statistics about the overall database.}
\label{tab:dbStats} 
\centering
\begin{tabular}{lr}
\hline
\hline
Class & Number \\
\hline
Jupiter Trojans                          &  80    \\
Resonant (3:2)                           &  47    \\
Resonant (others)                        &  28    \\
Long-period comets (LPCs)              &  14    \\
Short-period comets (SPCs)             &  136     \\
Centaurs                                 &  35   \\
Scattered disk objects (SDOs)            &  30  \\
Detached disk objects (DDOs)             &  28   \\
Cold classical disk objects (cold CDOs)  &  89  \\
Hot classical disk objects (hot CDOs)    &  68  \\
\hline
\hline
Database &\\
\hline
Objects&   555    \\ 
Epochs&    2045    \\ 
Papers&    100    \\ 
\hline
\end{tabular}
\end{table}
\begin{landscape}
\begin{table*}
\caption{Object class and colour data for some example objects. The
  full table is available in electronic form at the CDS, and
  maintained up-to-date on the MBOSS colour web site.}
\label{tab:allColours}
\centering
%
\begin{tabular}{lcccccccccc} 
\hline 
Object & Class$^{(1)}$&Epochs$^{(2)}$& M11 $\pm\sigma$& Grt$^{(3)}$ $\pm\sigma$& B-V $\pm\sigma$& V-R $\pm\sigma$& R-I $\pm\sigma$& V-J $\pm\sigma$& J-H $\pm\sigma$& H-K $\pm\sigma$ \\ 
\hline 
\hline 
1P/Halley               & SP Comet  &  2 &13.558 $\pm$ 0.596 & 4.266 $\pm$ 2.015 & 0.720 $\pm$ 0.040 & 0.410 $\pm$ 0.030 & 0.390 $\pm$ 0.060 &---  &---  &--- \\
9P/Tempel 1              & SP Comet  & 14 &14.560 $\pm$ 0.460 & 9.959 $\pm$ 0.826 &---  & 0.468 $\pm$ 0.010 & 0.469 $\pm$ 0.013 &---  &---  &--- \\
1172-Aneas            & J.Trojan  &  2 & 8.729 $\pm$ 0.042 & 9.261 $\pm$ 0.863 & 0.727 $\pm$ 0.030 & 0.510 $\pm$ 0.022 & 0.400 $\pm$ 0.030 & 1.577 $\pm$ 0.036 & 0.430 $\pm$ 0.042 & 0.135 $\pm$ 0.036\\
1994~ES$_{2}$           & Classic Cold &  1 & 7.509 $\pm$ 0.130 &---  &---  &---  &---  &---  &---  &--- \\
1997~SZ$_{10}$          & Res.(other) &  1 & 8.148 $\pm$ 0.060 &28.167 $\pm$ 2.909 & 1.140 $\pm$ 0.080 & 0.650 $\pm$ 0.030 &---  &---  &---  &--- \\
1999~CF$_{119}$         & Detached   &  3 & 6.933 $\pm$ 0.133 &14.704 $\pm$ 4.163 &---  & 0.622 $\pm$ 0.115 & 0.365 $\pm$ 0.106 &---  & 0.380 $\pm$ 0.226 &--- \\
1999~CX$_{131}$         & Res.(other) &  3 & 6.910 $\pm$ 0.107 &19.521 $\pm$ 4.293 & 0.918 $\pm$ 0.124 & 0.664 $\pm$ 0.128 & 0.434 $\pm$ 0.105 &---  & 0.370 $\pm$ 0.238 &--- \\
2003~HX$_{56}$          & Classic Hot &  1 & 7.030 $\pm$ 0.209 &-1.963 $\pm$13.676 &---  & 0.350 $\pm$ 0.226 & 0.260 $\pm$ 0.459 &---  &---  &--- \\
2060-Chiron           & Centaur   & 34 & 6.092 $\pm$ 0.069 & 0.114 $\pm$ 0.996 & 0.700 $\pm$ 0.020 & 0.361 $\pm$ 0.017 & 0.325 $\pm$ 0.023 & 1.199 $\pm$ 0.110 & 0.294 $\pm$ 0.079 & 0.065 $\pm$ 0.094\\
5145-Pholus           & Centaur   & 41 & 7.165 $\pm$ 0.076 &48.354 $\pm$ 1.930 & 1.261 $\pm$ 0.121 & 0.788 $\pm$ 0.036 & 0.822 $\pm$ 0.054 & 2.612 $\pm$ 0.048 & 0.391 $\pm$ 0.047 &-0.037 $\pm$ 0.047\\
12917-1998~TG$_{16}$   & J.Trojan  &  2 &11.388 $\pm$ 0.073 &11.839 $\pm$ 1.962 & 0.724 $\pm$ 0.042 & 0.537 $\pm$ 0.042 & 0.410 $\pm$ 0.069 & 1.707 $\pm$ 0.066 & 0.475 $\pm$ 0.086 & 0.205 $\pm$ 0.081\\
15760-1992~QB$_{1}$    & Classic Cold &  5 & 6.979 $\pm$ 0.094 &27.538 $\pm$ 4.806 & 0.869 $\pm$ 0.143 & 0.707 $\pm$ 0.093 & 0.651 $\pm$ 0.166 &---  &---  &--- \\
15809-1994~JS          & Res.(other) &  2 & 7.479 $\pm$ 0.160 &28.194 $\pm$ 6.000 &---  & 0.760 $\pm$ 0.170 & 0.480 $\pm$ 0.149 &---  & 0.460 $\pm$ 0.705 &--- \\
19308-1996~TO$_{66}$   & Classic Hot & 16 & 4.520 $\pm$ 0.042 & 2.077 $\pm$ 2.168 & 0.671 $\pm$ 0.057 & 0.389 $\pm$ 0.043 & 0.356 $\pm$ 0.053 & 0.997 $\pm$ 0.101 &---  &--- \\
20000-Varuna           & Classic Hot &  8 & 3.455 $\pm$ 0.090 &26.843 $\pm$ 1.917 & 0.906 $\pm$ 0.052 & 0.637 $\pm$ 0.040 & 0.628 $\pm$ 0.040 & 2.010 $\pm$ 0.050 & 0.564 $\pm$ 0.070 &-0.038 $\pm$ 0.105\\
20161-1996~TR$_{66}$   & Res.(other) &  1 &---  &---  &---  &---  &---  &---  & 0.470 $\pm$ 0.296 &--- \\
24835-1995~SM$_{55}$   & Classic Hot & 10 & 4.332 $\pm$ 0.040 & 0.272 $\pm$ 1.805 & 0.652 $\pm$ 0.032 & 0.357 $\pm$ 0.043 & 0.356 $\pm$ 0.052 & 1.010 $\pm$ 0.050 &-0.270 $\pm$ 0.163 &--- \\
35671-1998~SN$_{165}$  & Classic Cold &  6 & 5.679 $\pm$ 0.320 & 6.857 $\pm$ 3.068 & 0.712 $\pm$ 0.095 & 0.444 $\pm$ 0.078 & 0.437 $\pm$ 0.083 & 1.270 $\pm$ 0.050 &---  &--- \\
42355-Typhon           & Scattered  & 12 & 7.252 $\pm$ 0.054 &12.657 $\pm$ 1.354 & 0.758 $\pm$ 0.039 & 0.525 $\pm$ 0.022 & 0.414 $\pm$ 0.053 & 1.560 $\pm$ 0.045 & 0.406 $\pm$ 0.087 & 0.160 $\pm$ 0.071\\
42355-Typhon+B         & Scattered  &  1 &---  & 9.763 $\pm$ 0.000 &---  &---  &---  &---  &---  &--- \\
50000-Quaoar           & Classic Hot &  4 & 2.220 $\pm$ 0.029 &29.224 $\pm$ 1.706 & 0.958 $\pm$ 0.035 & 0.650 $\pm$ 0.020 & 0.610 $\pm$ 0.028 & 2.180 $\pm$ 0.058 & 0.360 $\pm$ 0.050 & 0.030 $\pm$ 0.057\\
52747-1998~HM$_{151}$  & Classic Cold &  1 & 7.417 $\pm$ 0.100 &24.891 $\pm$ 4.691 & 0.930 $\pm$ 0.090 & 0.620 $\pm$ 0.050 &---  &---  &---  &--- \\
90377-Sedna            & Detached   &  3 & 1.077 $\pm$ 0.065 &32.954 $\pm$ 2.972 & 1.131 $\pm$ 0.079 & 0.686 $\pm$ 0.077 & 0.657 $\pm$ 0.067 & 2.320 $\pm$ 0.060 & 0.290 $\pm$ 0.222 & 0.050 $\pm$ 0.314\\
90482-Orcus            & Res. 3:2   &  6 & 1.982 $\pm$ 0.099 & 2.761 $\pm$ 1.831 & 0.664 $\pm$ 0.041 & 0.370 $\pm$ 0.039 & 0.390 $\pm$ 0.045 & 1.070 $\pm$ 0.042 & 0.120 $\pm$ 0.051 & 0.053 $\pm$ 0.055\\
134340-Pluto            & Res. 3:2   &  4 &-0.881 $\pm$ 0.400 & 7.607 $\pm$ 0.681 & 0.867 $\pm$ 0.016 & 0.515 $\pm$ 0.035 & 0.400 $\pm$ 0.010 &---  &---  &--- \\
134340-Pluto+B          & Res. 3:2   &  1 &---  & 3.334 $\pm$ 0.000 & 0.710 $\pm$ 0.002 &---  &---  &---  &---  &--- \\
134340-Pluto+C          & Res. 3:2   &  1 &---  &-2.234 $\pm$ 0.000 & 0.644 $\pm$ 0.028 &---  &---  &---  &---  &--- \\
134340-Pluto+D          & Res. 3:2   &  1 &---  &18.074 $\pm$ 0.000 & 0.907 $\pm$ 0.031 &---  &---  &---  &---  &--- \\
136108-Haumea           & Classic Hot &  3 & 0.217 $\pm$ 0.030 &-0.010 $\pm$ 0.850 & 0.631 $\pm$ 0.025 & 0.370 $\pm$ 0.020 & 0.320 $\pm$ 0.020 & 1.051 $\pm$ 0.020 &-0.044 $\pm$ 0.037 &-0.111 $\pm$ 0.048\\
136199-Eris             & Detached   & 64 &-1.462 $\pm$ 0.036 & 3.866 $\pm$ 0.823 & 0.805 $\pm$ 0.015 & 0.389 $\pm$ 0.049 & 0.363 $\pm$ 0.061 & 0.849 $\pm$ 0.108 & 0.080 $\pm$ 0.072 &-0.280 $\pm$ 0.085\\
136472-Makemake         & Classic Hot &  1 &---  & 4.693 $\pm$ 0.900 & 0.828 $\pm$ 0.022 &---  &---  &---  &---  &--- \\
...\\
\hline
\end{tabular}

(1) Classes refer to Gladman et al, SSBN08. M11 is the absolute $R$ magnitude;
(2) Number of epochs;
(3) Grt is the spectral gradient $\cal S$.

\end{table*}
\end{landscape}
\begin{table}
\centering
\caption{For some objects, list of the references used, and number of
  epochs. The full table is available in electronic form at the CDS, and
  maintained up up-to-date on the MBOSS colour web site.}
\label{tab:photRef}
\begin{tabular}{lp{4cm}l}
\hline
Object & References & Nr.\\
\hline
\hline
              1994~ES$_{2}$
          &       \citet{Green+97}
 &   1\\
     7066-Nessus           
          &           \citet{D+98}
\\
          &           \citet{RT99}
\\
          &      \citet{Bauer+03b}
\\
          &       \citet{D98mboss}
\\
          &           \citet{TR98}
 &  17\\
    15760-1992~QB$_{1}$    
          &           \citet{JL01}
\\
          &           \citet{TR00}
\\
          &         \citet{HBO+01}
\\
          &         \citet{BEN+11}
\\
          &           \citet{R+97}
 &   5\\
    19308-1996~TO$_{66}$   
          &          \citet{JPH07}
\\
          &           \citet{JL98}
\\
          &           \citet{RT99}
\\
          &          \citet{ORH00}
\\
          &          \citet{GHL01}
\\
          &         \citet{DAV+00}
\\
          &           \citet{JL01}
\\
          &         \citet{HBO+01}
\\
          &          \citet{She10}
\\
          &         \citet{Bar+99}
\\
          &       \citet{D98mboss}
\\
          &           \citet{TR98}
 &  16\\
...\\
\hline
\end{tabular}

\end{table}
\begin{figure*}
\centering 
\includegraphics[width=8.5cm]{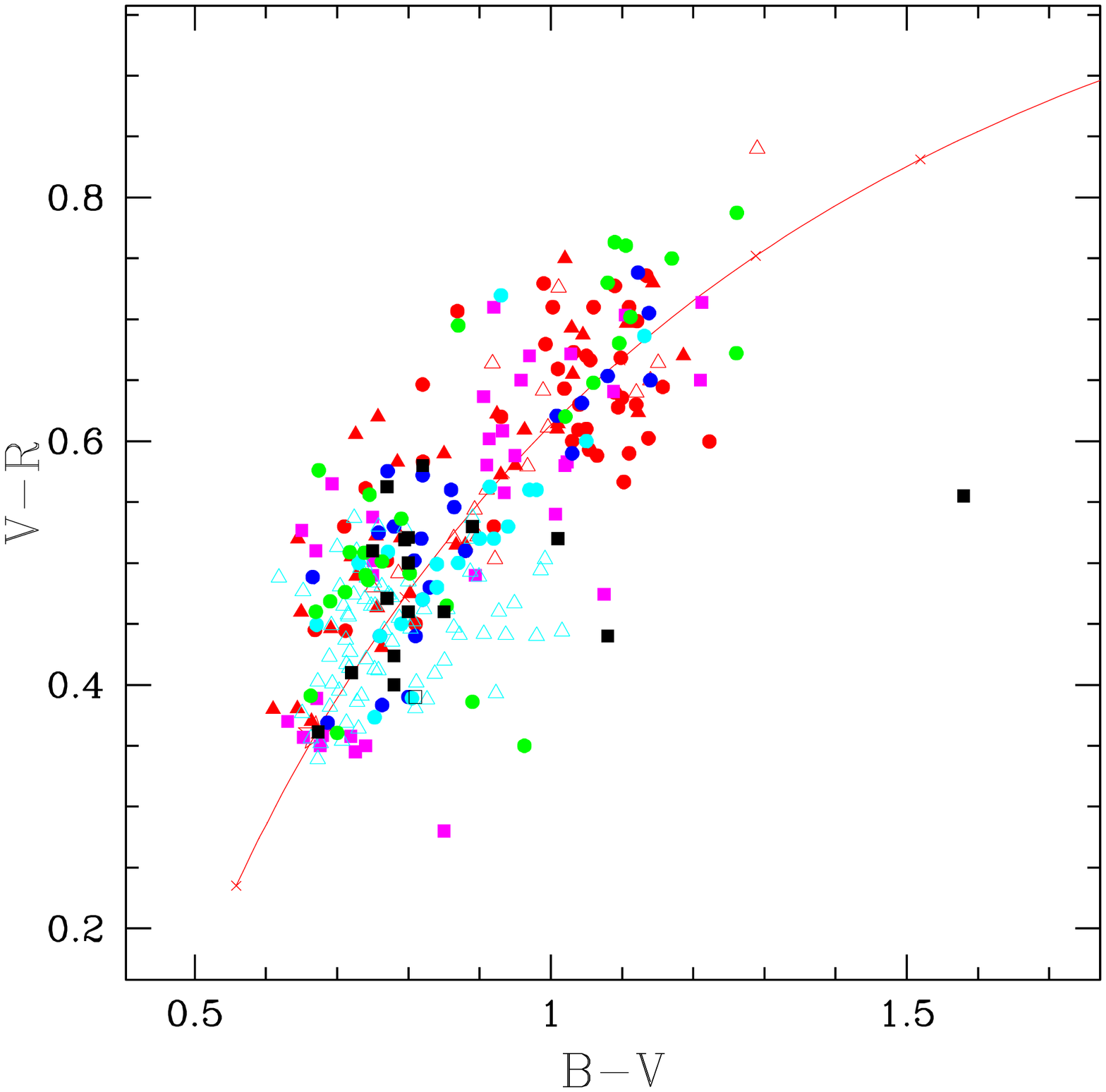}\hfill
\includegraphics[width=8.5cm]{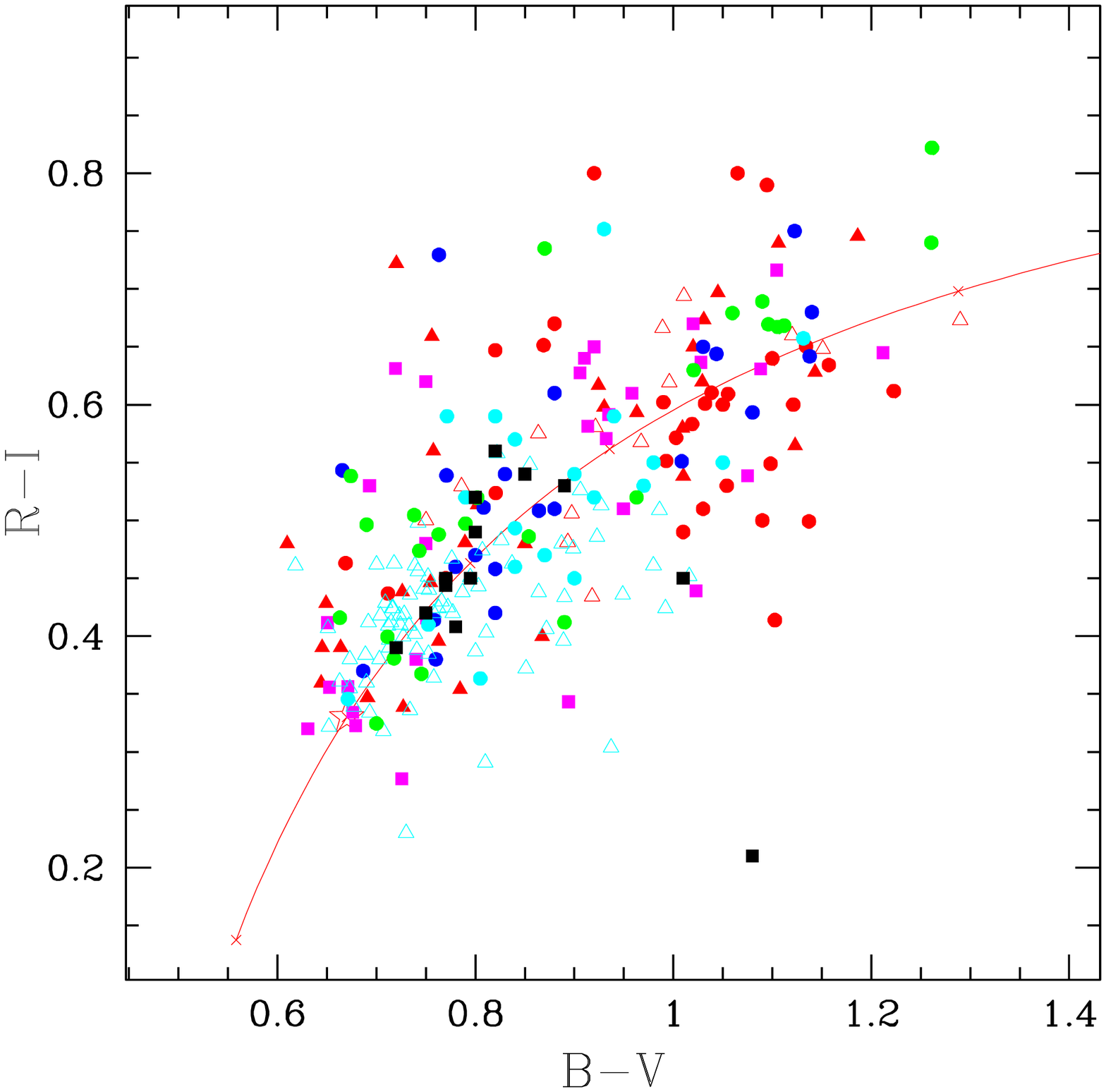}\\
\includegraphics[width=8.5cm]{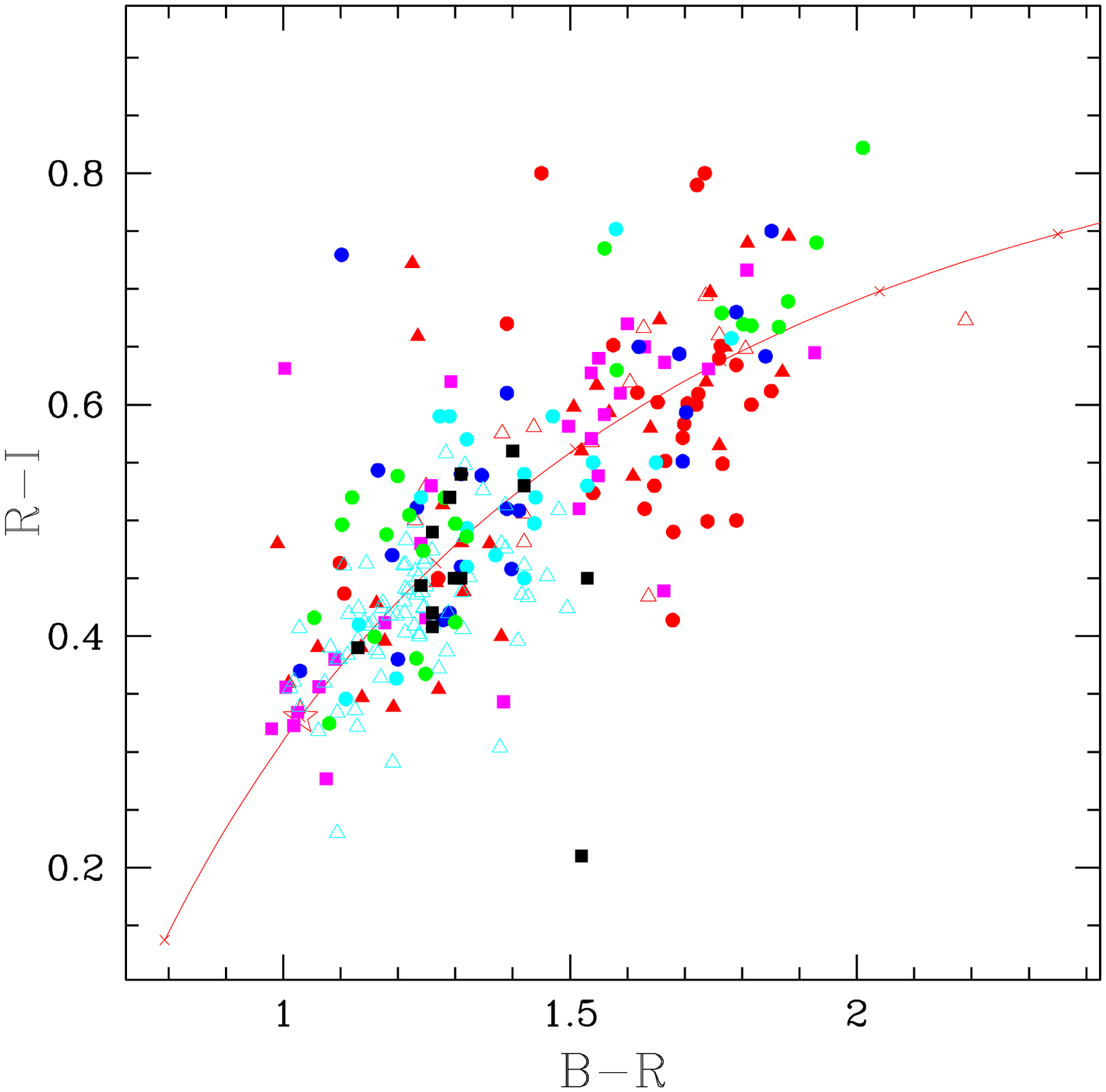}\hfill
\includegraphics[width=8.5cm]{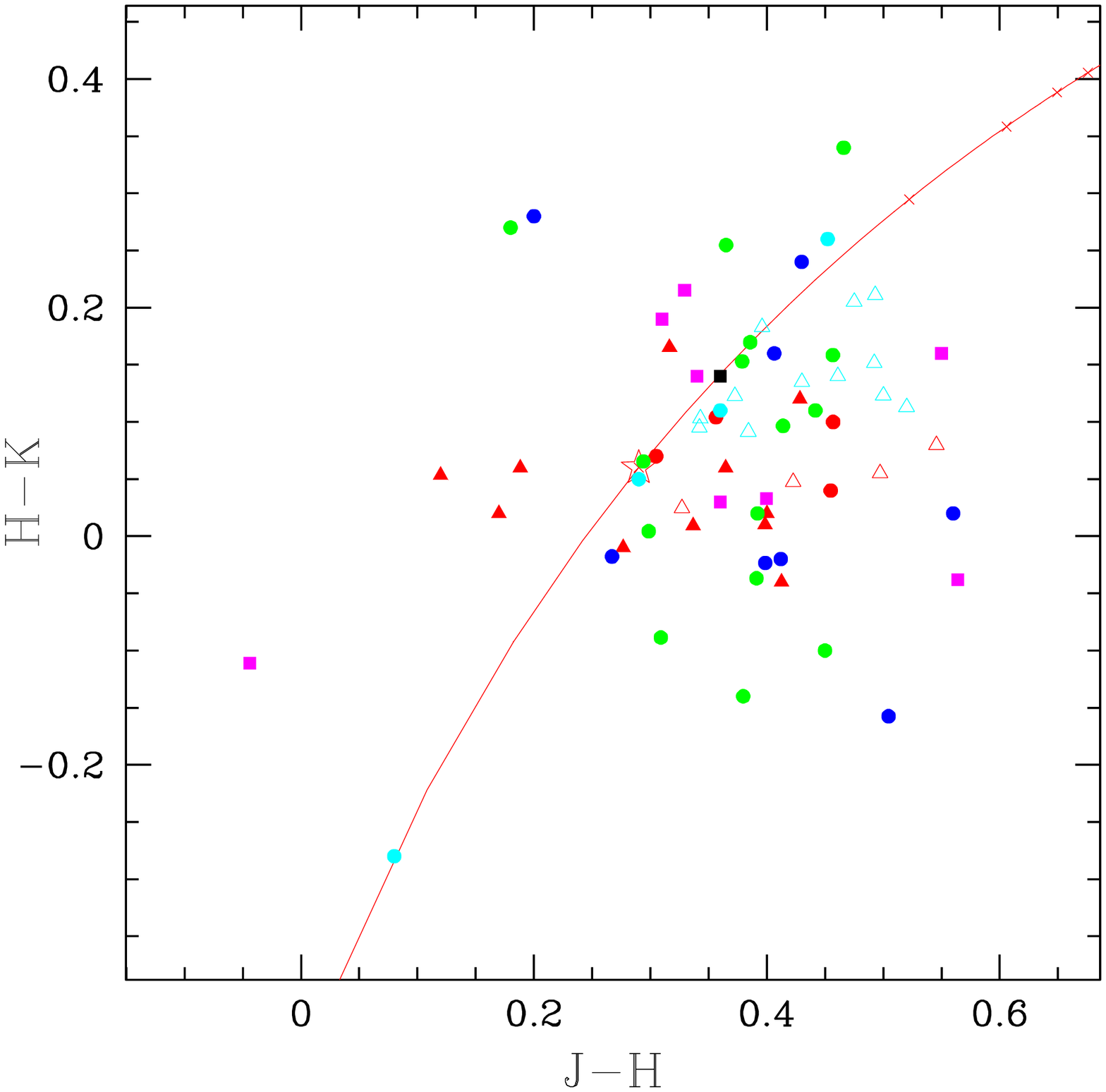}\\
\caption{Selected colour-colour diagrams of the objects. The
  physico-dynamical class of the objects is identified by their
  symbols, which are explained in Fig.~\ref{fig:legend}. The red star
  indicates the solar colours, and the red line is the locus of objects
  with a flat reflection spectrum, with a small mark every
  10\%/100nm. The other colour combinations are available from the MBOSS
  site.}
\label{fig:colcol}%
\end{figure*}
\begin{figure}
\centering 
\includegraphics[width=5cm]{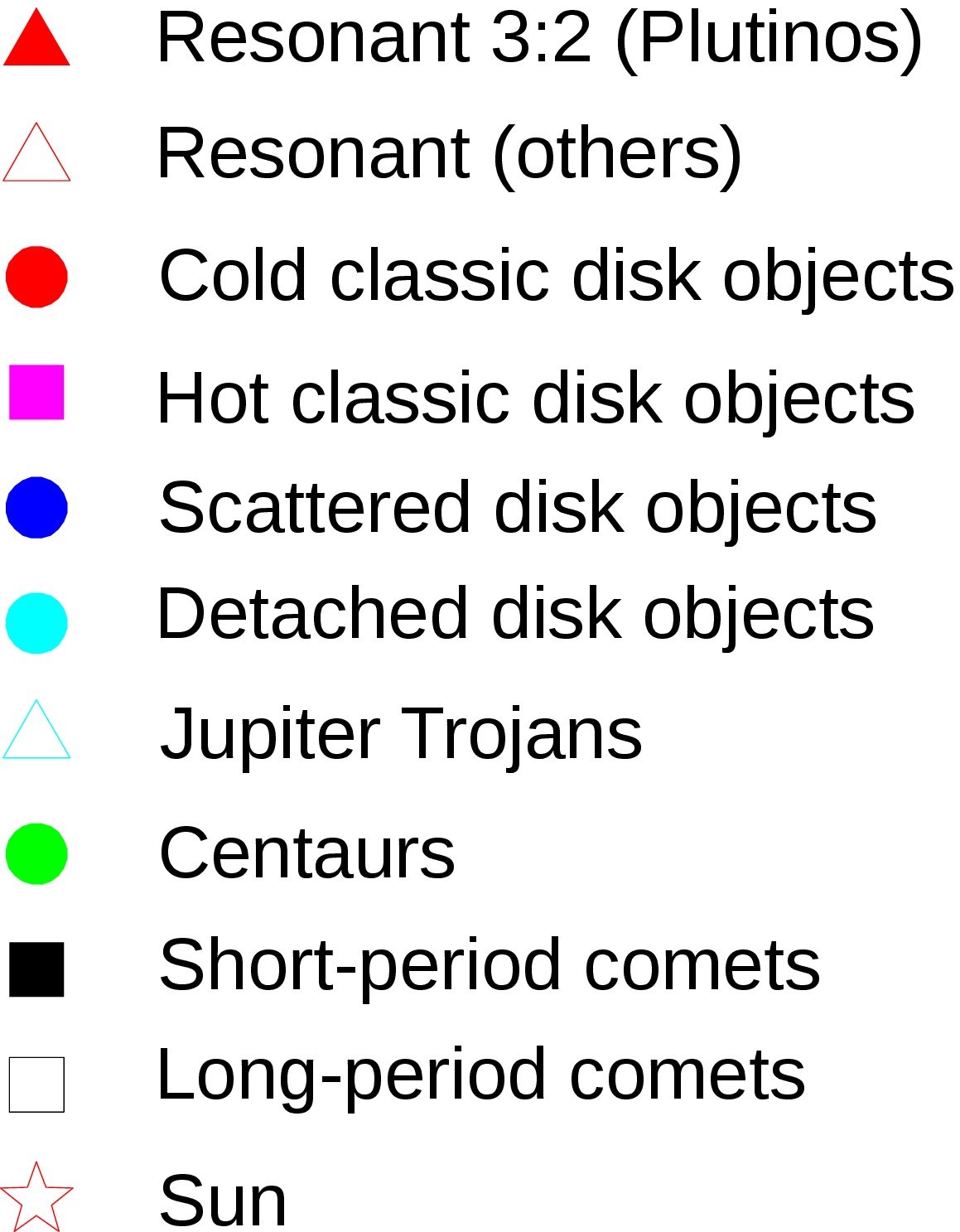}
\caption{Legend of the symbols used thorough this paper.}
\label{fig:legend}%
\end{figure}

\begin{figure}
\centering 
\includegraphics[width=5.7cm]{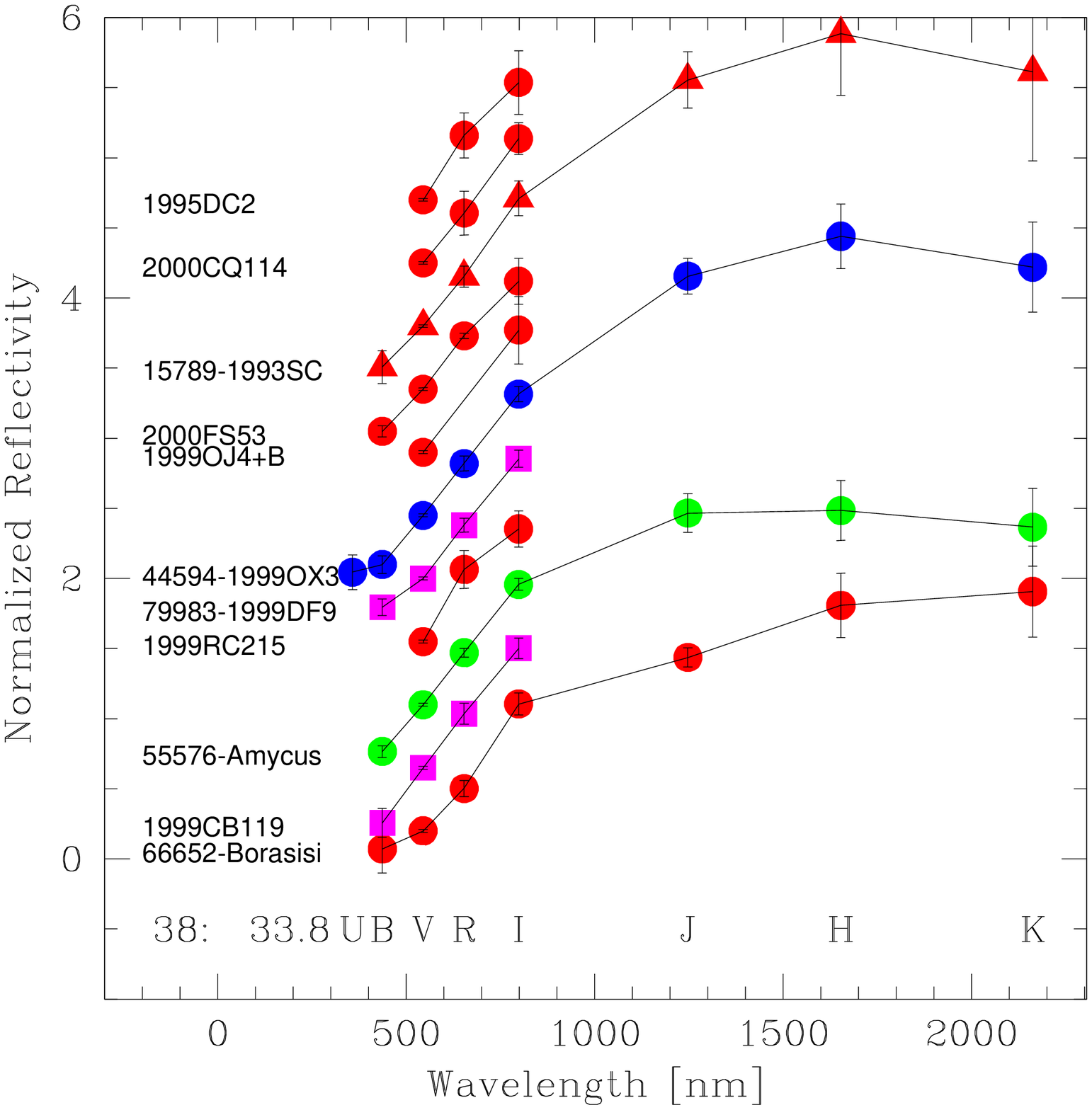}\hfill
\includegraphics[width=5.7cm]{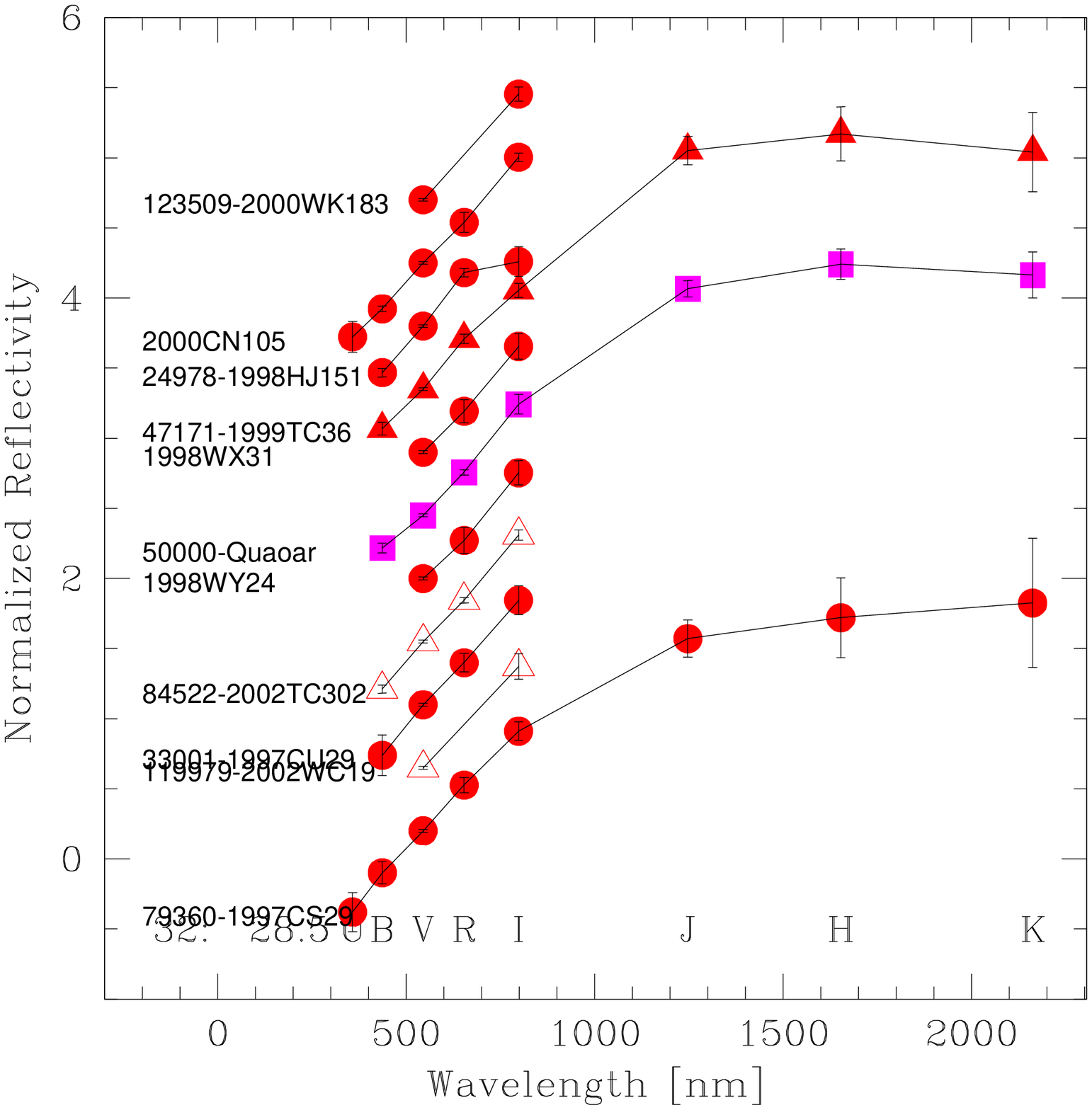}\\
\includegraphics[width=5.7cm]{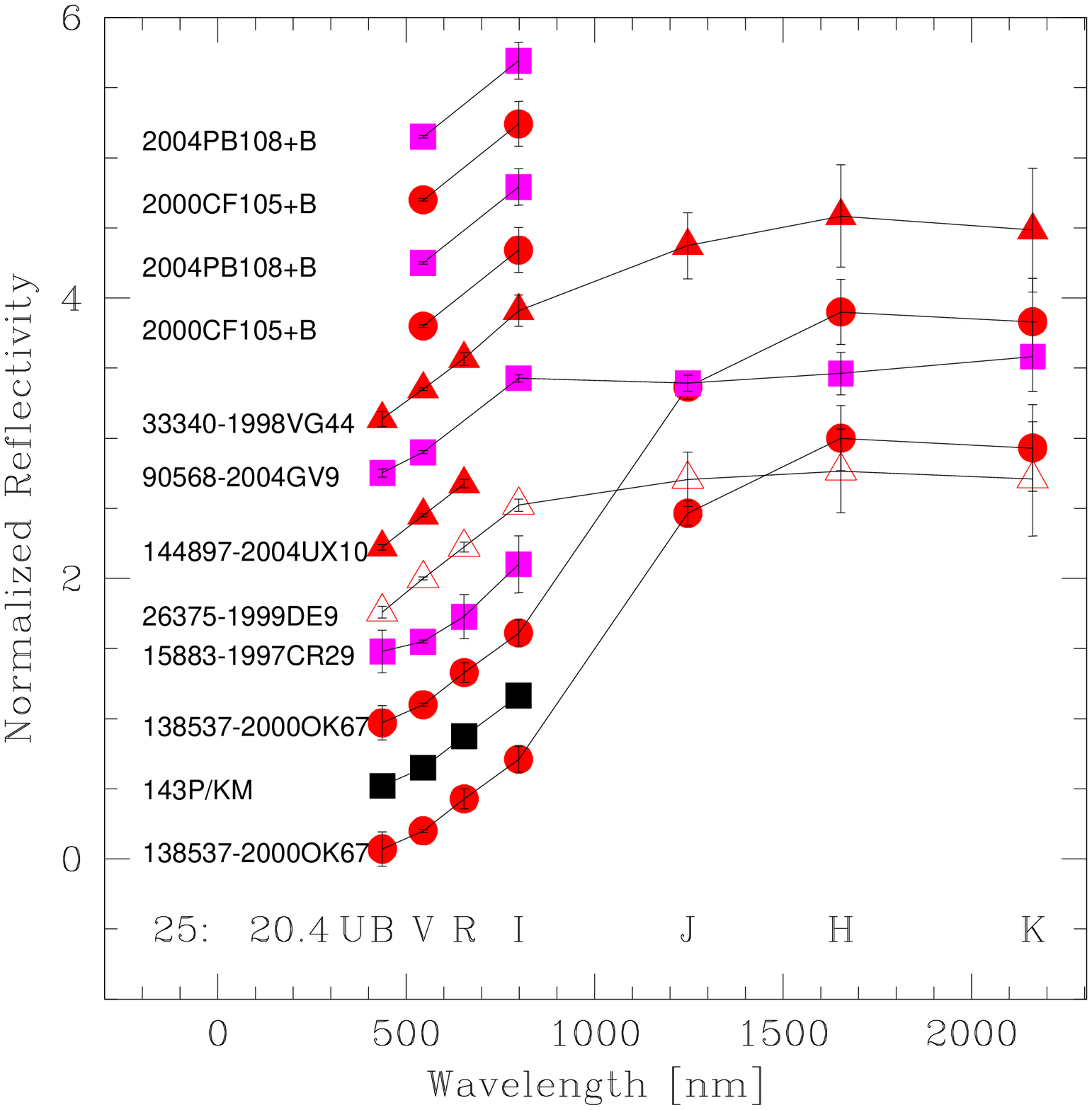}\hfill
\includegraphics[width=5.7cm]{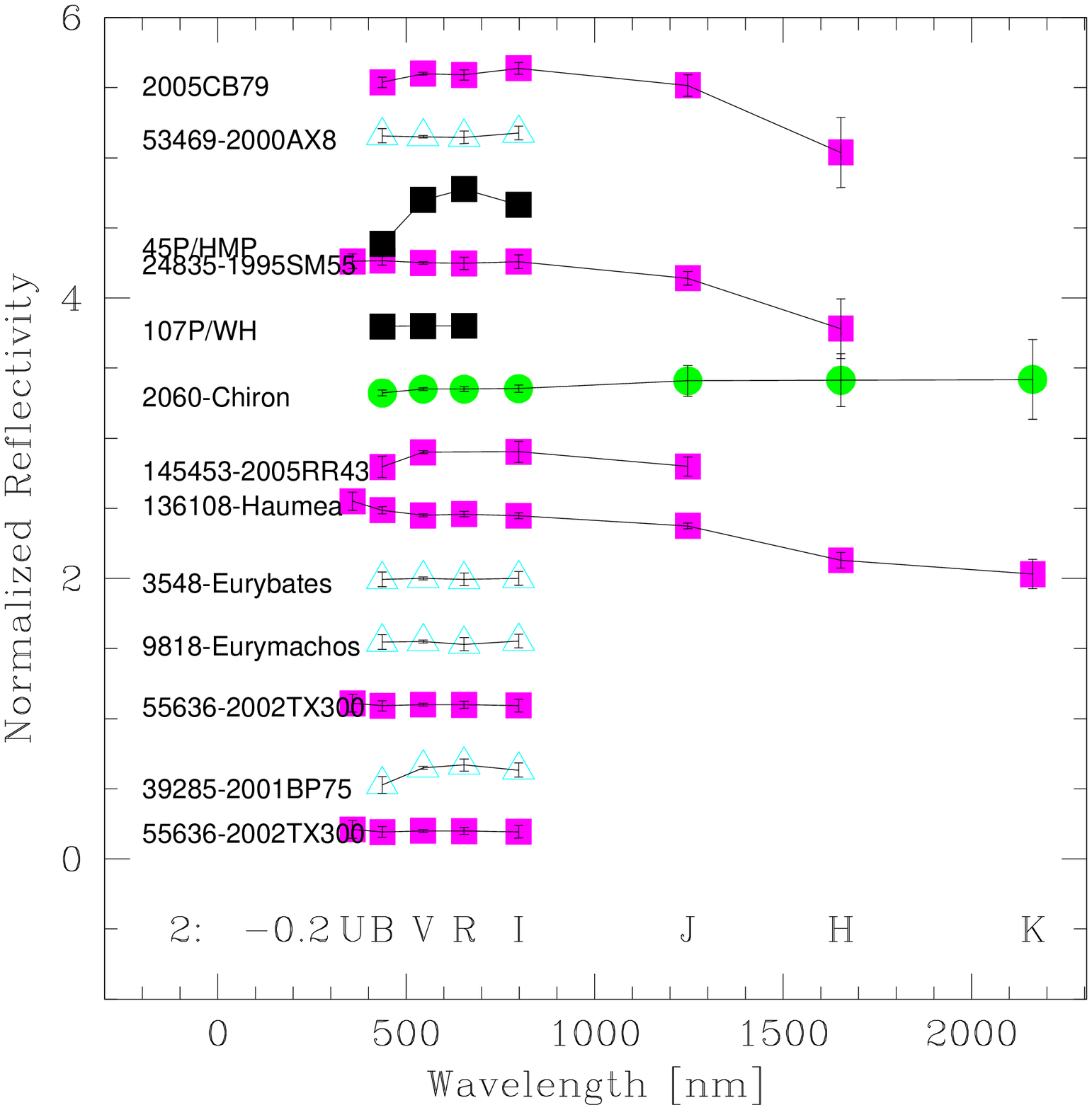}\\
\caption{Examples of reflectivity spectra for a set of objects. The
  reflectivity is normalized to unity for the $V$ filter, and the
  spectra are shifted for clarity. The physico-dynamical classes of 
  the objects are identified by their symbols, which are explained 
  in Fig.~\ref{fig:legend}. The spectra for all the objects are
  available online on the MBOSS site.}
\label{fig:spectra}%
\end{figure}



Table~\ref{tab:photRef} shows examples of the list of references per 
object that were included in this study; the information for all objects 
is given in the online supplement.


\label{sect:reddening} 

Figure~\ref{fig:colcol} presents some of the representative
colour-colour diagrams.  As a reference, the solar colours are
indicated by a red star (see paper~I, Table 2 for a list of the solar
colours and their references). The red ``reddening line'', introduced in
paper~I, marks the locus of objects with a perfectly linear
reflectivity spectrum over the considered colours, with a tick mark
every 10 units of $\cal S$. For colour diagrams in the visible range,
the distance from a data point to the reddening line therefore
indicates a ``bend'' in the reflectivity spectrum. The symbols used
indicate the dynamical class, and are explained in
Fig.~\ref{fig:legend}. Figure~\ref{fig:spectra} shows examples of the
coarse reflectivity spectra for a set of objects obtained from their
filter photometry. All the other plots are available at the MBOSS
site.


Figure~\ref{fig:orbitGrt} displays the spectral gradient $S$ as a
function of the main orbital elements, the orbital excitation, and the
$M(1,1)$ magnitude of the objects. Other similar plots, for all the
colours, are available online. The symbols indicate the dynamical
class.


Figure~\ref{fig:histogram} shows the histogram and the cumulative
distribution for an example colour and the spectral gradient. Other
similar plots are available on-line.


\section{Results from the statistical analysis of the enhanced MBOSS
  database}\label{REF-3}

\subsection{Statistical tools}\label{REF-3.1}


Comparing the distribution by comparing their histograms ``by eye'' is
unreliable: the size of the bins can cause artifacts or hide real
features, and what appears as a strong difference can actually be of
no significance. Similarly, the eye is a very powerful tool to detect
alignments and clustering, even when these are insignificant.

The colour, spectral gradient and $M(1,1)$
distributions of the various MBOSS populations were therefore compared
using a set of simple statistical tests that quantifies the
significance of the apparent differences, i.e. the $t$-test, the
$f$-test, and the Kolmogorov-Smirnov test. The tests are described in
detail in Appendix B of paper~I and references therein and can be
briefly summarized as follows:

\begin{itemize}
\item The Student $t$-test indicates whether the mean values of the two
  distributions are statistically different. The implementation used
  here deals properly with distributions with different variances.
\item The $f$-test considers whether two distributions have
  significantly different variances.
\item Finally, the Kolmogorov-Smirnov (KS) test uses all of the
  information contained in the distributions (and not just their means
  and variances) to estimate whether they are different.
\end{itemize}

For these tests to give meaningful results, the samples compared must
be sufficiently large. We set the threshold at 15 units for $t$- and
$f$-tests, and at 20 units for the KS test.

\subsection{The case of the brightest objects}
\label{REF-m11cutoff}

{ Large KBOs have atypical surface properties, e.g high albedo
  \citep{Sta+08,Bro08}, flatter spectral gradients, and CH$_4$
  ice-dominated spectra \citep{BTR05}.  For instance, Pluto is known
  to have a tenuous but captured atmosphere that can be re-deposited on
  the surface \citep{ST98, Pro+08}, while (136199)~Eris has an
  extremely high albedo \citep{Sic+11} and a spectrum dominated by
  CH$_4$ ice \citep{BTR05} and a nitrogen ice signature
  \citep{Dum+07}. \cite{Sha10} discusses the presence of volatiles
  (nitrogen, methane, and carbon monoxide) on bodies large enough to
  retain an atmosphere.  The transition from small objects with
  volatile-free surfaces (such as Charon or Quaoar) to large objects
  with volatile-rich ones (such as Pluto) can be explained with a
  simple model of atmospheric escape formulated by \citet{SB07}.}

We therefore decided to remove the largest objects (using their
intrinsic brightness as a proxy) for some studies, so that their
potentially different characteristics do not pollute the colour
distributions of the other, ``normal'' MBOSSes.  To select a
cut-off value, we estimated the radius of the objects whose escape
velocity $v_{\rm esc}$ is equal to the velocity of material ejection
via cometary activity $v_{\rm ej}$. The escape velocity is given by
\begin{equation}
v_{\rm esc} = \sqrt{ \frac{ 2 G M}{R} },
\end{equation}
where $G$ is the gravitational constant, $M$ the mass of the object,
and $R$ its radius. Assuming a density $\rho \sim 1000$kg~m$^{-1}$ to
be able to infer the mass from the volume of the object, 
\begin{equation}
v_{\rm esc} \sim 7.5 ~ 10^{-4}\, R
\end{equation}
in IS units. To estimate the velocity of material ejection, $v_{\rm
  ej}$ (in m~s$^{-1}$), we used the relation
\begin{equation}
v_{\rm ej} = 580 ~ r^{-0.5} \sqrt{\frac{\mu_0}{\mu}},
\end{equation}
where $r$ is the heliocentric distance in AU, and $\mu_o/\mu$ the
ratio of the molecular mass of the species driving the activity to
that of water. This relation was obtained by measuring the expansion
of cometary comae, and is supported by a theoretical analysis ---see
\cite{Del82} for a discussion. This velocity is the terminal velocity
of the gas in the case of a small comet, and it is directly controlled by
the thermal velocity of the sublimating gas. For the TNOs, we used
$\mu_o/\mu=0.64$ for CO. By equating $v_{\rm ej}$ to $v_{\rm esc}$, we
obtained a critical radius $R_c$ above which an object is likely to
retain some of the material ejected by cometary activity
\begin{equation}
where R_c \sim \frac{ 6.2 ~ 10^5 }{\sqrt{r}}.
\end{equation}
$R_c$ is in m and $r$ in AU. \cite{ADS+04} discuss the evolution of
$R_c$ with different species and distance. For our purpose, it is
enough to say that $R_c \sim 150$~km at $r=17$~AU and 100~km at 43~AU.
\cite{Sta+08}, based on their large set of measurements with the
Spitzer Space Telescope, indicate that large TNOs have higher than
average albedos. We used $p=0.2$ to convert $R_c$ in an absolute
magnitude, leading to 
\begin{equation}
M(1,1) \sim 1.9 + 2.5 \log{ r },
\end{equation}
again with $r$ in AU. This gives $M(1,1)\sim 5$ at
$r=17$~AU, and 6 at 43~AU. In what follows, we took
$M(1,1)=5$ as a conservative limit, below which objects are likely to
keep at least part of their atmospheres in the Centaur and TN regions,
thus be potentially affected by different resurfacing
processes.  This choice is partly arbitrary. We performed the
following analysis with different cut-off values, leading to similar
results.

To verify whether we were justified in separating the bright
$M(1,1)$ objects from the others, we compared their colours and
gradients with those of the faint objects. The Trojans were not
considered in this test, whose results are summarized in
Table~\ref{tab:testLarge}.

\begin{table}
\caption{Averge colours and dispersions of all MBOSSes (excluding
  Trojans), comparing 
  those with $M(1,1)<5$ and the others. }
\label{tab:testLarge}
\begin{center}
\begin{tabular}{crcrccc}
\hline
Colour  &  N  &  Aver./$\sigma$& N& Aver./$\sigma$&  $t$-Prob &$f$-Prob\\
  & \multicolumn{2}{c}{$M(1,1)<5$} & \multicolumn{2}{c}{$M(1,1) \ge 5$}\\
\hline
$J-H$& 22& 0.27$\pm$ 0.25& 84& 0.41$\pm$ 0.20&0.025&0.149\\
$H-K$& 16& 0.04$\pm$ 0.07& 31& 0.07$\pm$ 0.12&0.279&0.036\\
\hline
\end{tabular}
\end{center}

Notes:   The $q$ cut-off is set at the median value; N is the number of
   measurements; $t$-Prob and $f$-Prob are the probabilities that the
   two sub-samples are randomly extracted from the same distribution,
   as evaluated with the $t-$ and $f-$tests. The other colours show
   insignificant differences.
\end{table}

With the exception of the $J-H$ colour, the
bright and the faint objects have compatible mean colours and overall
distribution. In the infrared, the bright objects have marginally
bluer colours ($J-H=0.27\pm0.25$) than the faint ones
(($J-H=0.41\pm0.20$); this has only a probability of 2.5\% of occurring by
chance, i.e. this difference is marginally significant. Similarly, the
bright objects have a slightly broader distribution of
$H-K$ colour than the fainter ones (at the 3\% level). When performing
these tests on individual or groups of classes, the result was
a slightly more significant for the resonant objects, but could not
be tested for the other groups, which did not have enough bright
objects. Nevertheless, removing the non-resonant objects increased the
significance of the result for the resonant objects.
Additionally, one must consider the
possibility that the larger dispersion in $H-K$ for the fainter objects
is connected to the lower signal-to-noise ratio in $K$, as the sky is
much brighter in this band and the instruments tend to be less
sensitive there.
Considering the colours of the various dynamical classes in the
visible wavelength range for both the bright ($M(1,1)<5$) and faint objects,
the colour distributions of the bright objects were found to be
indistinguishable from those of the fainter ones, indicating that the
object populations might be rather uniform despite  some larger
objects possibly having some different surface properties (for instance, albedo)
and  intrinsic activity possibly playing a role in resurfacing the
bodies.
We note, however, that the NIR colours of individual KBOs can be
affected if a significant fraction of their surfaces is covered with
ices which display strong absorptions (such as $CH_4$), particularly
in $H$ and $K$ bands. For example, the colours of Pluto, Eris and 2005
$FY_9$ \citep[see][]{Bro08} show this effect.  The presence of these
ice absorption lines can be inferred from the unusual negative slope
of the NIR spectral gradients (i.e. from $J$ to $H$ and $K$
bands).

In summary, the infrared colours of the objects with bright absolute
magnitudes differ slightly  from those of the fainter
objects. We therefore  separated them for most colour tests,
but we  also re-incorporated them into some tests of the visible
colours where more objects were helpful.  

On the basis of these results, one might consider including the
intrinsically brighter objects in the remainder of the study. However,
we decided not to: if they were indeed similar, they would only
marginally increase the size of the sample, without changing our
conclusions. If they eventually turned out to be different, as
possibly suggested by the NIR colours, we did the right thing in
keeping them separate.

\subsection{Global characteristics of individual classes}            
The MBOSS database that we analyse contains photometric measurements
of ten different dynamical classes (see Fig. ~\ref{fig:legend}), i.e.
the SPCs, the LPCs, the Jupiter Trojans, Centaurs, Kuiper Belt objects
in 3:2 (Plutinos = Res. 3:2) and in other orbital resonances
(Res.others) with Neptune, the classical disk objects (CDOs) in two
flavors of dynamically `hot' and `cold' CDOs, the scattered disk
objects (SDOs) and the detached disk objects (DDOs).

\begin{figure*}
   \centering 
   \includegraphics[width=8cm]{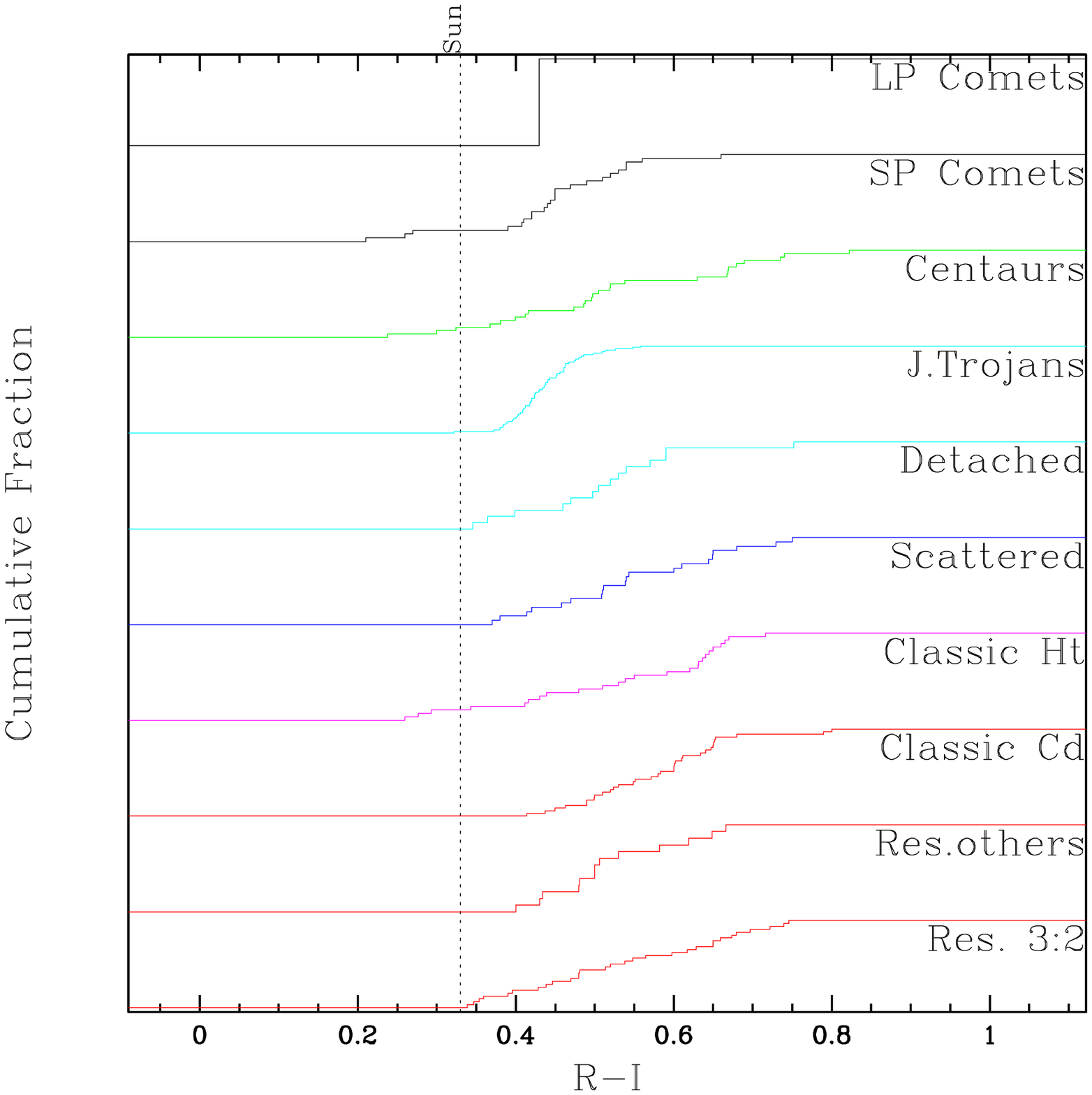}\hfill
   \includegraphics[width=8cm]{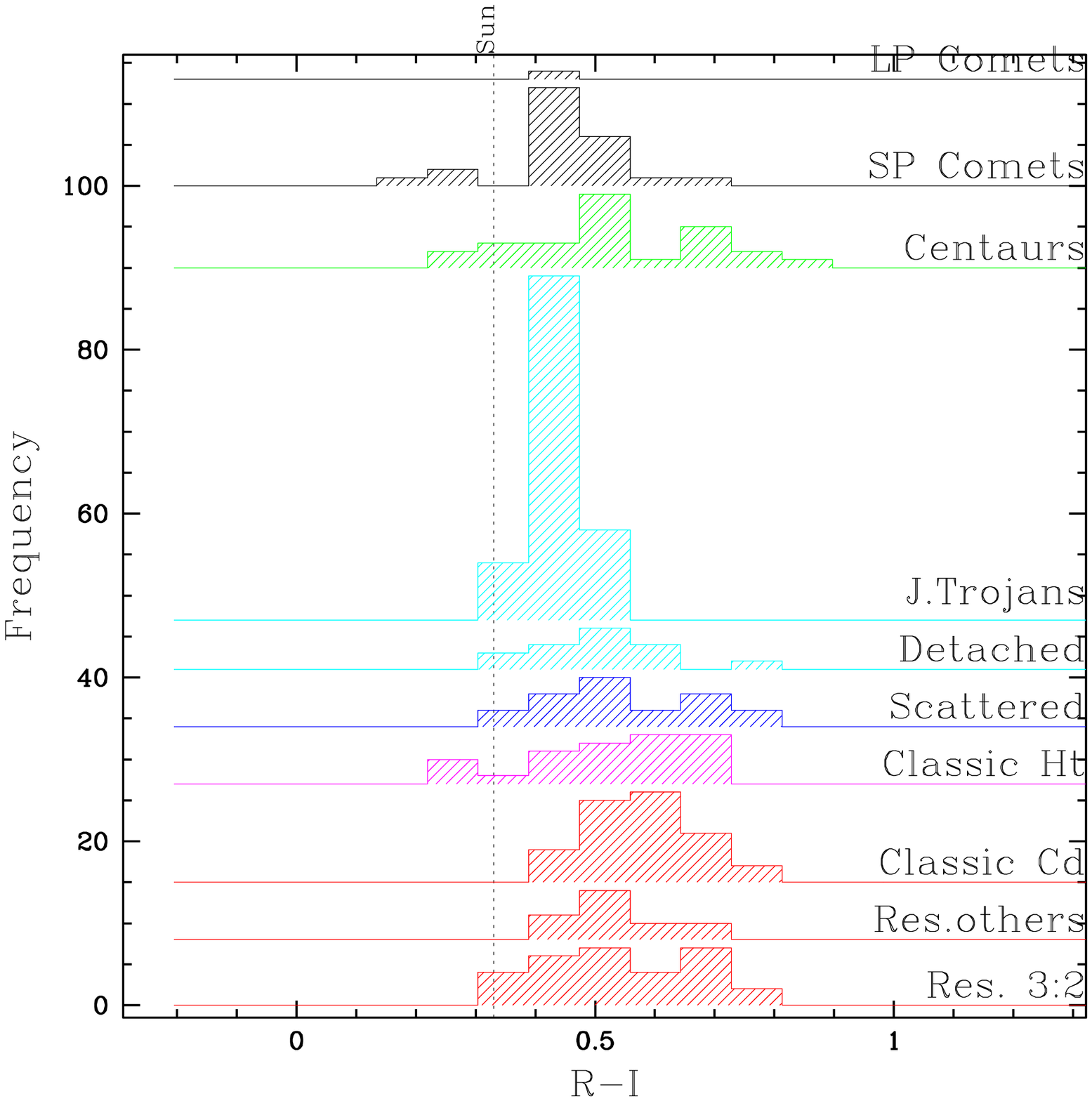}\\
   \includegraphics[width=8cm]{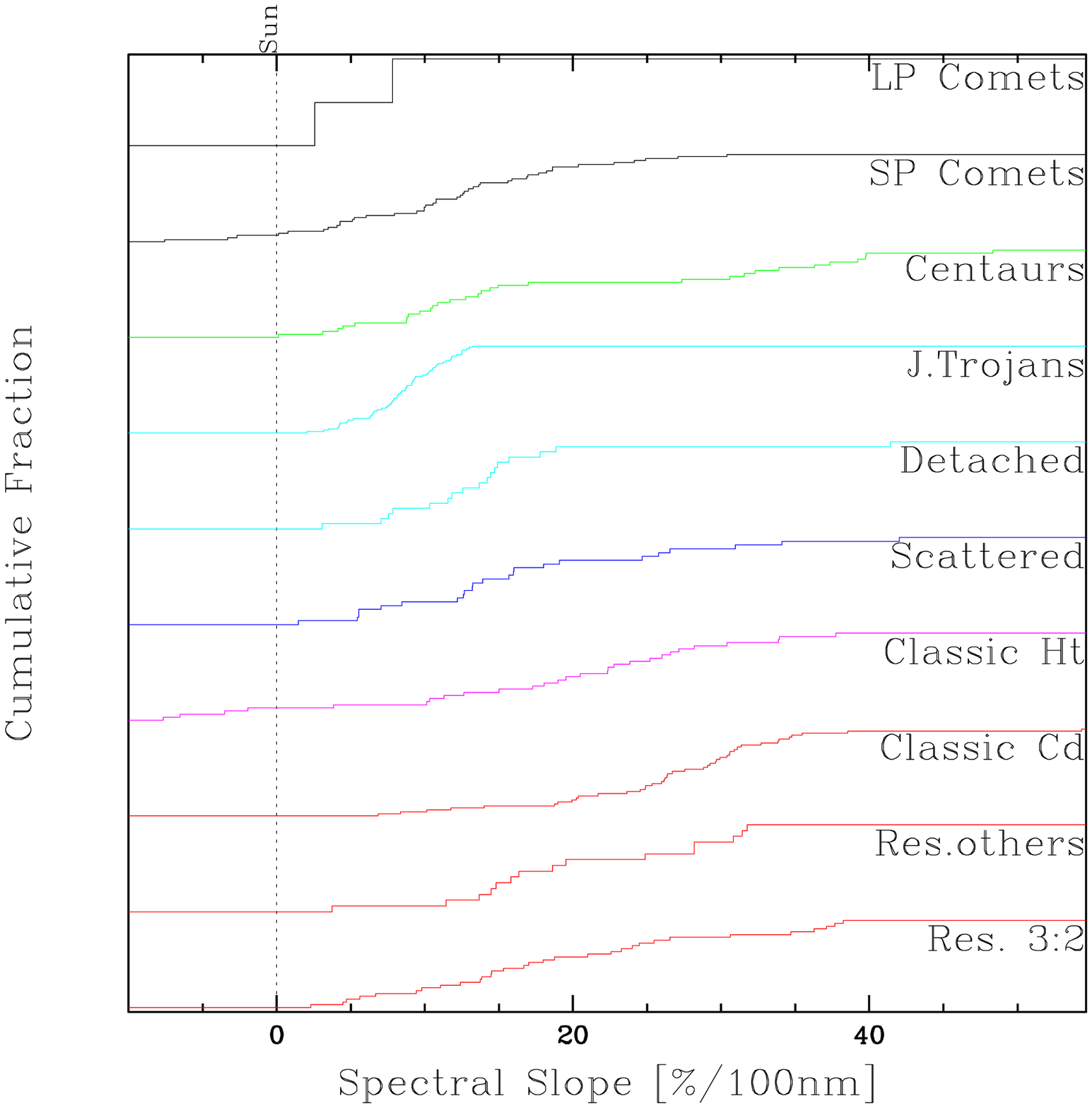}\hfill
   \includegraphics[width=8cm]{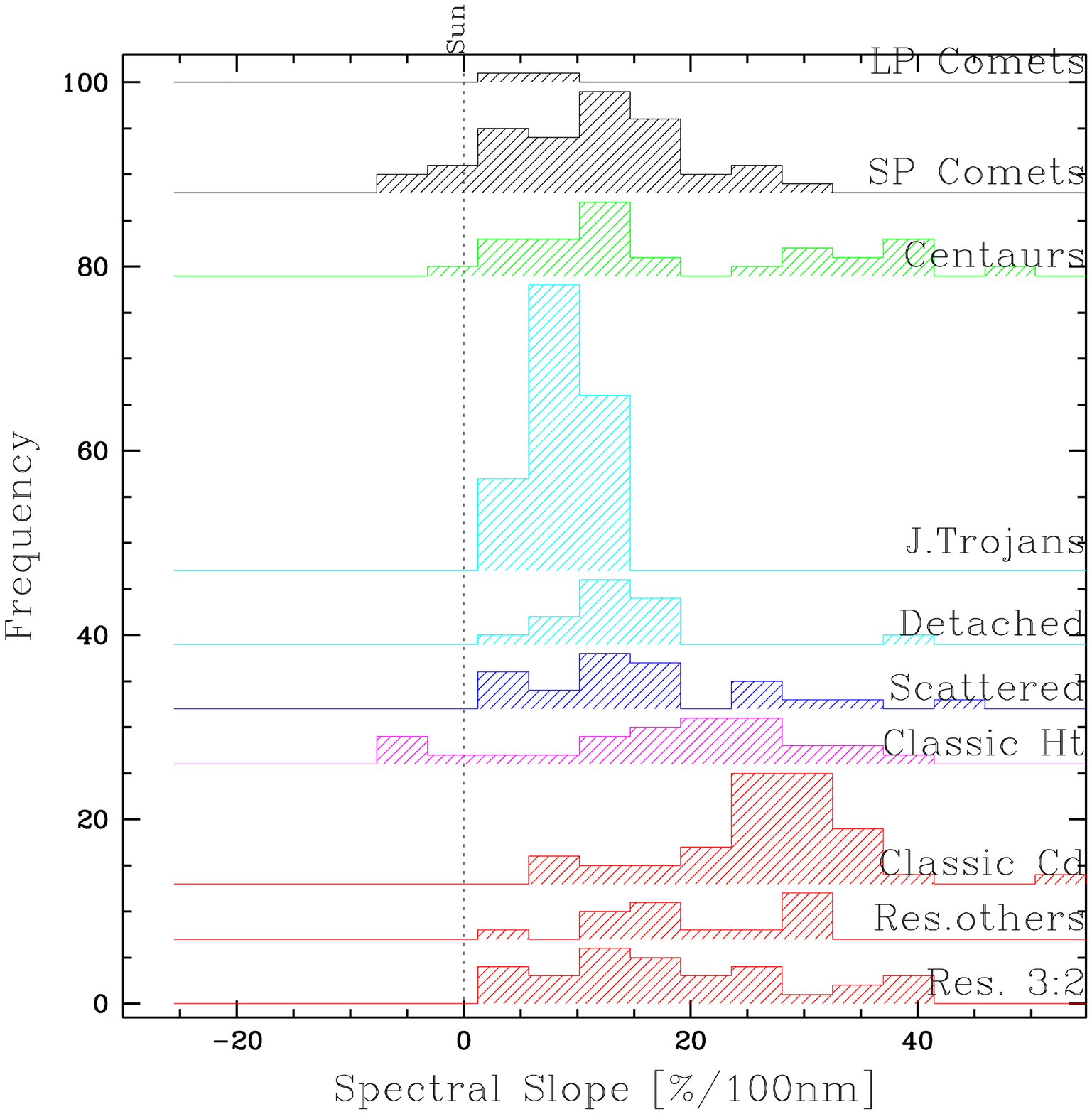}\\
   \caption{$R-I$ (top) and spectral gradient (bottom) cumulative
     distributions (left) and histogram (right). Solar value is
     indicated by a dotted line.  Similar plots are
     available on-line for the other colours.}
   \label{fig:histogram}%
\end{figure*}

Selected colour-colour diagrams of MBOSSes are shown in
Fig.~\ref{fig:colcol}. Colour distributions of the various classes of
objects are displayed in Fig.~\ref{fig:histogram}, for a selection of
colours and spectral gradients $\cal S$. At first sight, the ranges of
visible colours are similar for the several dynamical groups (Plutinos
and resonants, CDOs, SDOs, and Centaurs), ranging from slightly bluish
(-10\%/100~nm) compared to the Sun to very red (55\%/100~nm). Jupiter
Trojans do not appear to contain any very red objects (i.e. with
spectral gradients $> 20$\%/100~nm) as the other groups do. For LPCs,
the number of measured objects (14) may be too small to provide
representative results for the total population, although we note that
the available data indicate that the spectral gradients are between
about 0 and 10 \%/100nm, in close agreement with the other groups.

The mean colours (and corresponding variances) for the various
dynamical classes are listed in Table~\ref{tab:averClass}, top panel,
using all data in the MBOSS2 database, and the values for the
restricted data-set of objects with $M(1,1) \ge 5$~mag are in
Table~\ref{tab:averClass}, bottom panel. Many of these means are based on
only a small number of objects and
should be interpreted with care. 

\begin{figure}
a \includegraphics[width=8cm]{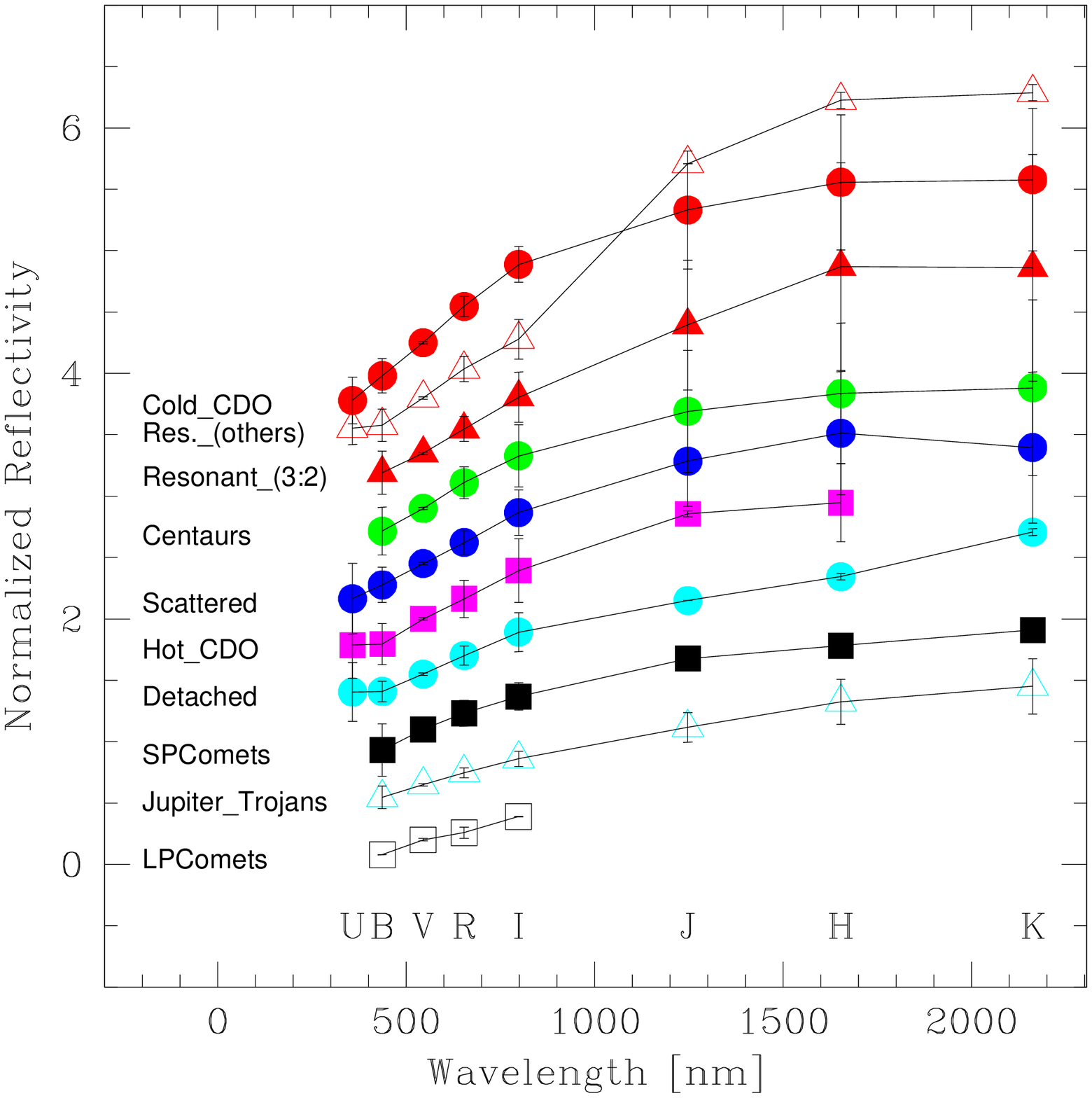}
b \includegraphics[width=8cm]{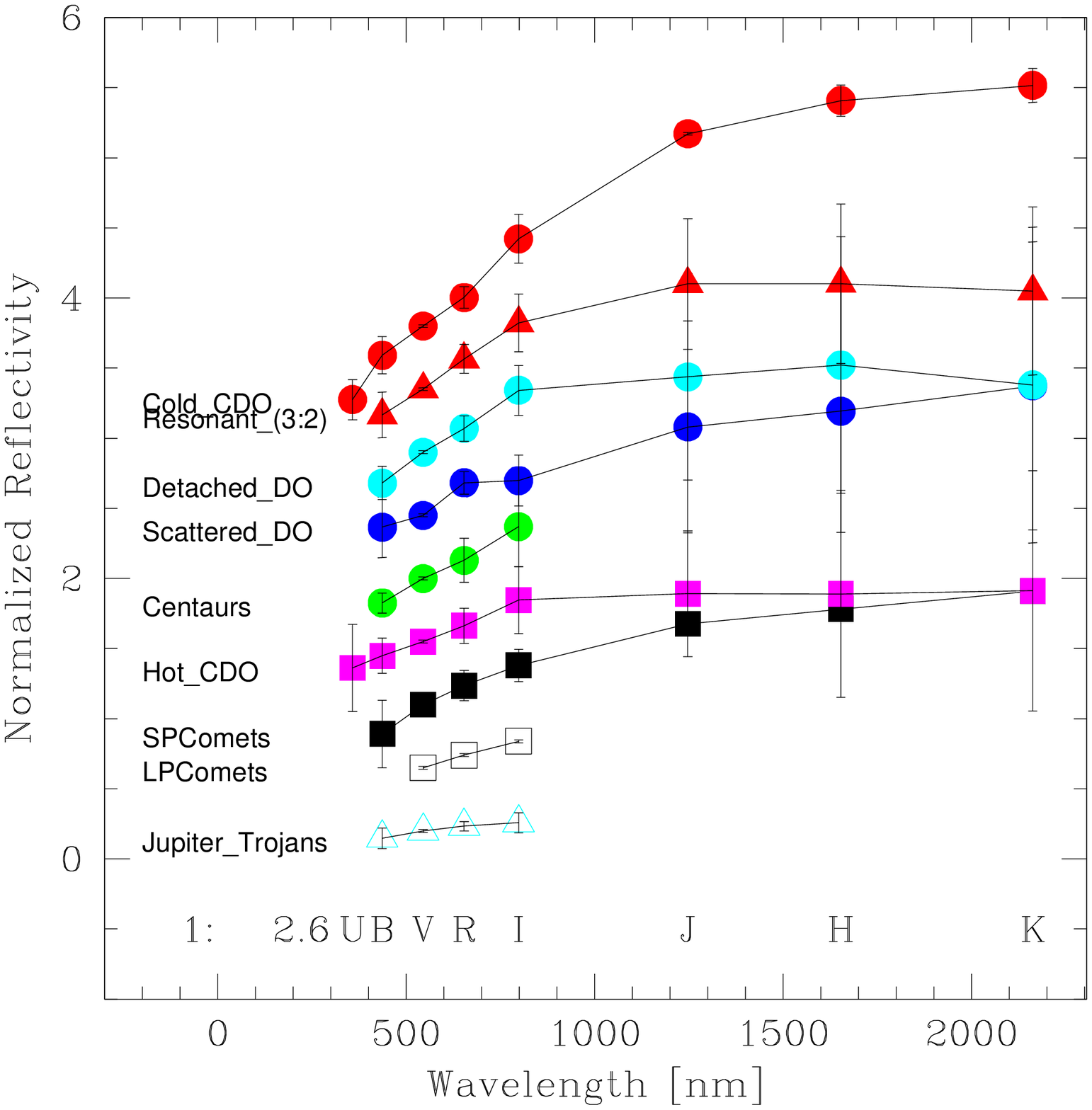}\\
c \includegraphics[width=8cm]{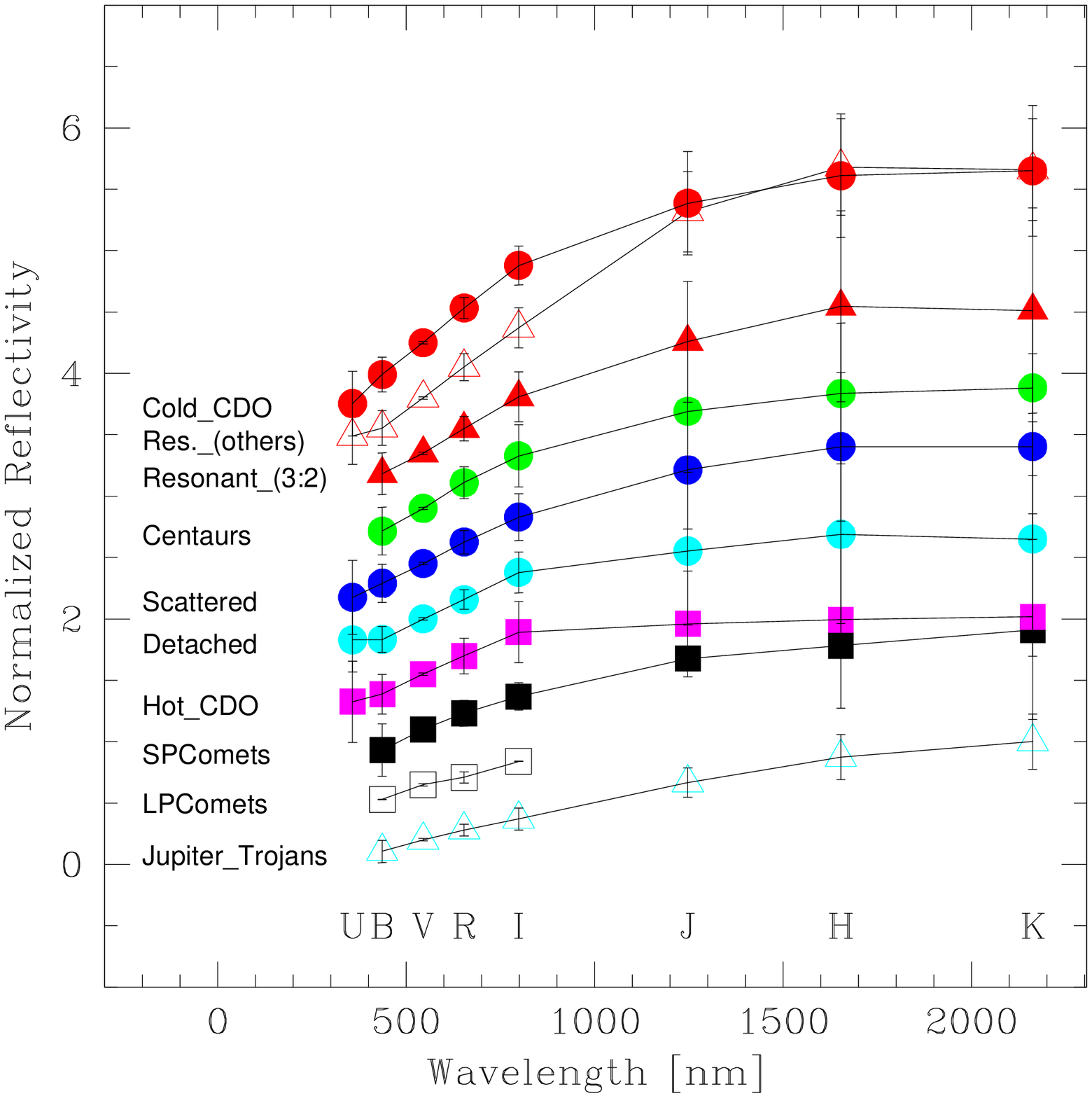}\\
 \caption[Average spectra]{\label{fig:averSpec} Average reflectivity
   spectra for the different physico-dynamical classes. a: small
   objects ($M(1,1)>5$ mag), b: large objects only ($M(1,1)<5$ mag),
   c: all objects}
\end{figure}

The average characteristics of the dynamical classes are also
displayed as reflectivity spectra in Fig.~\ref{fig:averSpec}. The red
slope in the visible wavelength range, although different for
individual groups, levels off in the NIR with transition between
the $I$ and $J$ or $H$ bands.

\subsection{Comparisons between classes}


In our comparison, we considered the photometric properties of
the dynamical classes all groups except LPCs, since are too
sparse to expect firm conclusions.

\subsubsection{Colours and spectral gradients}

    \begin{table*}
    \caption{Statistical comparisons of the spectral gradient
      distributions of pairs of MBOSS populations: $t$-test, $f$-test,
      and KS test.} 
    \label{tab:clasCompatibility}
    {\centering
\begin{tabular}{lrc||lllllllll}                                                                              
\hline
             &      &      &  Resonant    &  Classical   &  Classical    &  Scattered  &  Detached    &   Centaurs   &  Short        &  Long         &    Trojans  \\
             &      &      &  (others)    &  cold        &   hot         &  disk       &  disk        &              &  period       &  period       &      \\
             &      &      &              &              &               &  objects    &  objects     &              &  comets       &  comets       &      \\
\hline\hline
N  &      &      &           15 &           43 &           28  &          23 &           17 &           26 &            44 &             0 &           60\\
\hline
Resonant     &   31 & $t$  &        0.701 &        0.002 &        0.713  &       0.340 &        0.056 &        0.768 &        0.006  &               &        0.000\\
3:2          &      & $f$  &        0.367 &        0.209 &        0.505  &       0.777 &        0.264 &        0.163 &        0.334  &               &        0.000\\
             &      &  KS  &        0.810 &        0.000 &        0.817  &       0.612 &        0.023 &        0.438 &        0.072  &               &        0.000\\
             &      &      &Indist.&{Not comp.}&Indist.&Indist.&  Marginal    &Indist.&Marginal &  -/-   &{Not comp.}\\
\hline
Resonant     &   15 & $t$  &              &        0.019 &        0.485  &       0.211 &        0.034 &        0.968 &        0.005  &               &        0.000\\
others       &      & $f$  &       .      &        0.986 &        0.166  &       0.523 &        0.881 &        0.057 &        0.810  &               &        0.000\\
             &      &  KS  &              &        0.019 &        0.680  &       0.183 &        0.009 &        0.101 &        0.020  &               &        0.000\\
             &      &      &              &  Marginal    &Indist.&Indist.&Marginal&Indist.&Marginal    &  -/-          &{Not comp.}\\
\hline
Classical    &   43 & $t$  &              &              &        0.002  &       0.000 &        0.000 &        0.038 &        0.000  &               &        0.000\\
Cold         &      & $f$  &              &       .      &        0.051  &       0.402 &        0.839 &        0.006 &        0.747  &               &        0.000\\
             &      &  KS  &              &              &        0.006  &       0.000 &        0.000 &        0.000 &        0.000  &               &        0.000\\
             &      &      &              &              &               &{Not comp.}&{Not comp.}&{Not comp.}&{Not comp.}& -/- &{Not comp.}\\
\hline
Classic      &   28 & $t$  &              &              &               &       0.602 &        0.160 &        0.560 &        0.039  &               &        0.000\\
Hot          &      & $f$  &              &              &       .       &       0.377 &        0.106 &        0.471 &        0.094  &               &        0.000\\
             &      &  KS  &              &              &               &       0.161 &        0.002 &        0.284 &        0.040  &               &        0.000\\
             &      &      &              &              &               &Indist.& Marginal&Indist.&Marginal     &  -/-          &{Not comp.}\\
\hline
 Scattered   &   23 & $t$  &              &              &               &             &        0.355 &        0.286 &        0.114  &               &        0.001\\
Disk         &      & $f$  &              &              &               &      .      &        0.406 &        0.125 &        0.569  &               &        0.000\\
Objects      &      &  KS  &              &              &               &             &        0.280 &        0.358 &        0.359  &               &        0.000\\
             &      &      &              &              &               &             &Indist.&Indist.&Indist.&  -/-    &{Not comp.}\\
\hline
 Detached    &   17 & $t$  &              &              &               &             &              &        0.064 &        0.583  &               &        0.020\\
Disk         &      & $f$  &              &              &               &             &       .      &        0.031 &        0.662  &               &        0.000\\
Objects      &      &  KS  &              &              &               &             &              &        0.182 &        0.444  &               &        0.000\\
             &      &      &              &              &               &             &              &Indist.&Indist. &  -/-       &{Not comp.}\\
\hline
 Centaurs    &   26 & $t$  &              &              &               &             &              &              &        0.015  &               &        0.000\\
             &      & $f$  &              &              &               &             &              &       .      &        0.014  &               &        0.000\\
             &      &  KS  &              &              &               &             &              &              &        0.035  &               &        0.000\\
             &      &      &              &              &               &             &              &              &  Marginal     &   -/-         &{Not comp.}\\
\hline
 Short       &   44 & $t$  &              &              &               &             &              &              &               &               &        0.090\\
 Period      &      & $f$  &              &              &               &             &              &       .      &               &               &        0.000\\
 Comets      &      &  KS  &              &              &               &             &              &              &               &               &        0.000\\
             &      &      &              &              &               &             &              &              &               &   -/-         &{Not comp.}\\
\hline
\end{tabular}
 } 

Notes: The three tests are the Student $t-$ and $f-$tests and the KS
test. The number N indicates how many objects have measurements. The
result is the probability that the two distributions are randomly
extracted from the same population. The label indicates whether the
populations are significantly different (i.e., not compatible),
marginally different, or statistically indistinguishable (see text for
details).
    \end{table*}
\begin{table*}
\caption{Statistical comparisons of the $J-H$ colour distributions of
  pairs of MBOSS populations:  $t$-test, $f$-test, and
  KS test.
}
    \label{tab:clasCompatibilityIR}
\begin{center}
\begin{tabular}{lcc|lll}
\hline
         &    &   &Classic& Classic & Centaurs \\
         &    &   & cold  & hot     &   \\
\hline\hline
         &N   &   & 28    &  34     & 17  \\
\hline
Resonant & 21 & t & 0.498 & 0.119 & 0.308 \\
3:2      &    & f & 0.000 & 0.961 & 0.000 \\
         &    &KS & 0.449 & 0.677 & 0.984 \\
         &    &   &Not comp.&Indist.&Not comp.\\
\hline
Classic  & 28 & t &       & 0.121 & 0.332 \\
cold     &    & f &   -   & 0.000 & 0.703 \\
         &    &KS &       & 0.710 & 0.284 \\
         &    &    &      &Not comp.&Indist.\\
\hline
Classic  & 34 & t &       &       & 0.271 \\
hot      &    & f &   -   &   -   & 0.000 \\
         &    &KS &       &       & 0.499 \\
         &    &   &       &       &Not comp.\\
\hline
\end{tabular}
\end{center}

    Notes: The three tests are the Student $t-$ and $f-$tests and the
    KS test. The N number indicates how many objects have
    measurements. The classes that are not listed do not have
    sufficient measurements for these tests to be performed. 
    The result is the probability that the two distributions
    are randomly extracted from the same population. The label indicates
    whether the populations are significantly different, marginally
    different, or statistically indistinguishable (see text for
    details). 
\end{table*}


Comparing the colour distributions of the different physico-dynamical
classes, we can make the following observations (see
Fig.~\ref{fig:histogram}):
\begin{itemize}

\item The distributions of both the colours and the spectral gradients
  in the visible wavelength range differ among the groups, in terms of
  their extent, shape and peak location. Taking the spectral gradients
  as an example, the peak level increases starting with the Jupiter
  Trojans (5-10\%/100~nm), and continuing to the SPCs, Centaurs, DDOs,
  SDOs, Plutinos (and possibly the resonant objects) (10-20\%/100~nm),
  and the hot CDOs (20-30\%/100~nm), before ending with the cold CDOs
  which are the
  reddest objects (25-35\%/100~nm) among the MBOSSes. The colour ranges
  and distribution width can clearly be inferred from the cumulative
  distribution functions (see left panels of
  Fig. \ref{fig:histogram}).

\item Relevant secondary peaks in the frequency distributions may
  exist for Centaurs and Plutinos at higher reddening. At least for
  Centaurs, statistical arguments for a bimodal surface colour
  distribution have been presented in numerous papers \citep[eg][]{TR98,
    TR03, PEI+03, ADS+06, Teg+08}.  \cite{Teg+08} presented a detailed
  analysis of the Centaurs' bimodality, based on a sample of 26
  objects, i.e. almost as large as the one presented here (29 objects,
  so we do not present a new analysis). They concluded that the $B-R$
  distribution is bimodal with a confidence level of 99.5\%, with 10
  red Centaurs and 16 gray ones. The red ones have marginally smaller
  orbital inclinations, and higher albedos at the 99\% confidence
  level. 

\item The frequency distribution in $J-H$ instead gives a rather
  uniform picture for all dynamical groups with respect to colour
  range and peak position (with a few singular exceptions for Plutinos
  and CDOs). 
\end{itemize}

This global characterization indicates that parameters of the visible
spectral energy distribution are better diagnostics of both the global
surface reflectivity and differences among MBOSSes than the
NIR colours. However, we note that the NIR spectral
data of MBOSSes is more sensitive to 
compositional differences than visible one, because stronger
absorption bands, in particular of icy compounds, are found in the
NIR wavelength domain \citep{Bar+08}.

\begin{landscape}
\begin{table*}
\caption{Mean colours and variances, for the MBOSS populations: all objects} 
\label{tab:averClass}
\begin{tabular}{crrrrrrrrrr}
\hline
\hline 
colour& Res. 3:2  & Res.others& Classic cold& Classic hot& Scattered &
Detached  & J.Trojans & Centaurs  & SPCs & LPCs \\
\hline
\multicolumn{10}{l}{Full database}\\
\hline
$U-B$ &   0&   2&   3&  11&   7&   2&   0&   0&   0&   0\\
   &--- ---&   0.273$\pm$0.090&   0.593$\pm$0.123&   0.258$\pm$0.168&   0.334$\pm$0.144&   0.180$\pm$0.156&--- ---&--- ---&--- ---&--- ---\\
$B-V$ &  38&  16&  41&  38&  27&  22&  74&  28&  17&   1\\
   &   0.868$\pm$0.170&   0.975$\pm$0.142&   0.996$\pm$0.140&   0.864$\pm$0.163&   0.858$\pm$0.153&   0.869$\pm$0.109&   0.777$\pm$0.091&   0.892$\pm$0.195&   0.863$\pm$0.209&   0.810$\pm$0.000\\
$V-R$ &  38&  21&  49&  41&  25&  23&  80&  30&  43&   2\\
   &   0.558$\pm$0.101&   0.603$\pm$0.109&   0.630$\pm$0.086&   0.510$\pm$0.145&   0.536$\pm$0.096&   0.519$\pm$0.083&   0.445$\pm$0.048&   0.567$\pm$0.131&   0.494$\pm$0.105&   0.422$\pm$0.045\\
$V-I$ &  40&  23&  73&  51&  28&  24&  80&  26&  22&   1\\
   &   1.101$\pm$0.202&   1.181$\pm$0.164&   1.220$\pm$0.159&   1.010$\pm$0.249&   1.037$\pm$0.189&   1.039$\pm$0.170&   0.861$\pm$0.090&   1.077$\pm$0.252&   0.947$\pm$0.112&   0.878$\pm$0.000\\
$V-J$ &  16&   5&   5&  13&   9&   4&  12&  10&   1&   0\\
   &   1.838$\pm$0.493&   2.138$\pm$0.327&   1.960$\pm$0.421&   1.509$\pm$0.431&   1.753$\pm$0.481&   1.614$\pm$0.601&   1.551$\pm$0.120&   1.769$\pm$0.498&   1.630$\pm$0.000&--- ---\\
$R-I$ &  36&  19&  36&  35&  22&  21&  80&  26&  23&   1\\
   &   0.536$\pm$0.122&   0.554$\pm$0.090&   0.588$\pm$0.095&   0.506$\pm$0.139&   0.544$\pm$0.108&   0.515$\pm$0.099&   0.416$\pm$0.057&   0.525$\pm$0.151&   0.447$\pm$0.101&   0.430$\pm$0.000\\
$J-H$ &  21&  13&  28&  34&  13&  10&  12&  17&   1&   0\\
   &   0.442$\pm$0.285&   0.437$\pm$0.065&   0.398$\pm$0.082&   0.316$\pm$0.290&   0.400$\pm$0.124&   0.381$\pm$0.126&   0.434$\pm$0.064&   0.375$\pm$0.075&   0.360$\pm$0.000&--- ---\\
$H-K$ &  11&   4&   4&   8&   8&   4&  12&  15&   1&   0\\
   &   0.043$\pm$0.059&   0.052$\pm$0.023&   0.079$\pm$0.030&   0.077$\pm$0.117&   0.060$\pm$0.151&   0.035$\pm$0.228&   0.139$\pm$0.041&   0.085$\pm$0.143&   0.140$\pm$0.000&--- ---\\
\hline
Grt &  45&  25&  82&  56&  29&  26&  80&  30&  43&   2\\
   &  18.633$\pm$10.659&  23.471$\pm$9.362&  25.515$\pm$9.217&  15.387$\pm$11.887&  15.258$\pm$9.974&  15.459$\pm$8.873&   7.024$\pm$3.786&  19.320$\pm$13.842&  11.483$\pm$8.307&   5.188$\pm$3.692\\
\hline
$M(1,1)$ &  38&  21&  46&  45&  24&  23&  60&  29& 134&  13\\
   &   6.680$\pm$1.981&   6.235$\pm$1.393&   6.600$\pm$0.614&   5.928$\pm$1.703&   7.168$\pm$1.225&   5.718$\pm$2.181&  11.356$\pm$1.057&   9.456$\pm$1.654&  16.131$\pm$1.667&  13.550$\pm$3.601\\
\hline
\end{tabular}
\end{table*}
\end{landscape}

\begin{landscape}
\addtocounter{table}{-1}
\begin{table*}
\caption{Mean colours and variances, for the MBOSS populations,
  continued: Objects with $M(1,1) > 5 $mag} 
\begin{tabular}{crrrrrrrrrr}
\hline
\hline 
colour& Res. 3:2  & Res.others& Classic cold& Classic hot& Scattered &
Detached  & J.Trojans & Centaurs  & SPCs & LPCs \\
\hline
\multicolumn{10}{l}{Objects with $M(1,1) \ge 5$ mag}\\
\hline
$U-B$ &   0&   1&   2&   3&   7&   2&   0&   0&   0&   0\\
   &--- ---&   0.210$\pm$0.000&   0.525$\pm$0.049&   0.187$\pm$0.103&   0.334$\pm$0.144&   0.180$\pm$0.156&--- ---&--- ---&--- ---&--- ---\\
$B-V$ &  26&  10&  34&  22&  23&  15&  60&  28&  16&   1\\
   &   0.858$\pm$0.175&   0.943$\pm$0.131&   1.011$\pm$0.139&   0.919$\pm$0.167&   0.876$\pm$0.143&   0.836$\pm$0.084&   0.789$\pm$0.092&   0.892$\pm$0.195&   0.872$\pm$0.213&   0.810$\pm$0.000\\
$V-R$ &  31&  15&  42&  29&  23&  16&  60&  30&  43&   2\\
   &   0.555$\pm$0.102&   0.590$\pm$0.102&   0.642$\pm$0.082&   0.523$\pm$0.152&   0.532$\pm$0.098&   0.513$\pm$0.077&   0.461$\pm$0.040&   0.567$\pm$0.131&   0.494$\pm$0.105&   0.422$\pm$0.045\\
$V-I$ &  31&  13&  37&  25&  22&  15&  60&  26&  22&   1\\
   &   1.098$\pm$0.205&   1.114$\pm$0.162&   1.226$\pm$0.145&   1.050$\pm$0.258&   1.068$\pm$0.186&   1.010$\pm$0.159&   0.898$\pm$0.062&   1.077$\pm$0.252&   0.947$\pm$0.112&   0.878$\pm$0.000\\
$V-J$ &   9&   1&   4&   2&   6&   1&  12&  10&   1&   0\\
   &   1.912$\pm$0.528&   2.295$\pm$0.000&   1.932$\pm$0.480&   1.807$\pm$0.024&   1.796$\pm$0.369&   1.646$\pm$0.000&   1.551$\pm$0.120&   1.769$\pm$0.498&   1.630$\pm$0.000&--- ---\\
$R-I$ &  30&  13&  33&  25&  20&  14&  60&  26&  23&   1\\
   &   0.538$\pm$0.127&   0.521$\pm$0.084&   0.578$\pm$0.090&   0.529$\pm$0.137&   0.549$\pm$0.113&   0.510$\pm$0.104&   0.437$\pm$0.044&   0.525$\pm$0.151&   0.447$\pm$0.101&   0.430$\pm$0.000\\
$J-H$ &  14&   5&  17&  17&   9&   5&  12&  17&   1&   0\\
   &   0.518$\pm$0.319&   0.467$\pm$0.064&   0.401$\pm$0.071&   0.342$\pm$0.296&   0.417$\pm$0.133&   0.414$\pm$0.026&   0.434$\pm$0.064&   0.375$\pm$0.075&   0.360$\pm$0.000&--- ---\\
$H-K$ &   6&   1&   3&   0&   5&   1&  12&  15&   1&   0\\
   &   0.056$\pm$0.076&   0.080$\pm$0.000&   0.070$\pm$0.030&--- ---&  -0.004$\pm$0.114&   0.260$\pm$0.000&   0.139$\pm$0.041&   0.085$\pm$0.143&   0.140$\pm$0.000&--- ---\\
\hline
Grt &  31&  15&  43&  28&  23&  17&  60&  30&  43&   2\\
   &  19.051$\pm$10.558&  20.240$\pm$8.496&  26.488$\pm$8.567&  17.795$\pm$12.258&  16.526$\pm$10.121&  13.971$\pm$8.172&   8.538$\pm$2.759&  19.320$\pm$13.842&  11.483$\pm$8.307&   5.188$\pm$3.692\\
\hline
$M(1,1)$ &  32&  15&  45&  34&  23&  17&  60&  29& 134&  13\\
   &   7.392$\pm$0.802&   6.974$\pm$0.795&   6.635$\pm$0.571&   6.717$\pm$0.833&   7.275$\pm$1.124&   6.746$\pm$0.572&  11.356$\pm$1.057&   9.456$\pm$1.654&  16.131$\pm$1.667&  13.550$\pm$3.601\\
\hline
\end{tabular}

Notes: The number of objects included in each average is indicated.
Grt is the spectral gradient $\cal S$.
$M(1,1)$ is the $R$ absolute magnitude.
\end{table*}
\end{landscape}

We now compare the populations using the statistical tests described
in section \ref{REF-3.1}.  First of all, we stress that a
common feature of all the statistical tests applied is the estimate of a
probability that the two populations compared are {\em not} randomly
extracted from the same population. In other words, the statistical
tests can very firmly establish whether two populations  significantly
differ. A contrario, no statistical test can prove that two
populations are identical. For instance, the colour distributions of
the Centaurs and Jupiter Trojans (see
Tab.~\ref{tab:clasCompatibility}) are radically different with a very
high significance level: it is totally improbable that these two
populations were randomly extracted from the same reservoir. However,
the spectral gradient distributions of SDOs and DDOs indicate that these
objects could have been randomly extracted from a same pool, even
though this may be just a coincidence. 

Table~\ref{tab:clasCompatibility} summarizes the results of the
comparison between the spectral gradient $\cal S$ distribution of the
MBOSS populations using these tests. Following the discussion in
Section~\ref{REF-m11cutoff}, the tests were restricted to the objects
with $M(1,1) > 5$~mag. For each pair, the table lists the
probabilities that the two classes were randomly extracted from a
common population, based on the mean spectral gradient ($t$-test), the
spectral gradient variance ($f$-test), and the overall spectral
gradient distribution (KS test). Low probabilities indicate that the
two considered distributions are incompatible. The pairs for which
at least one of the tests returns a probability $< 0.001$ were flagged
as ``not compatible'', indicating they are significantly
different. Those where a test indicates a probability $<0.05$ are
marked as “marginally different”, suggesting that there is a
difference, but that it is not very strongly significant. The others
are statistically indistinguishable using these tests, which  again does
not mean that the objects are similar, just that these tests cannot
prove they are different. Similar tests were performed on the
individual colours, with similar results. We discuss here only the
spectral gradients, as they encompass most of the information present
in the visible spectrum range. 

In the infrared, the tests are possible only in $J-H$, and only some of the
classes have enough measurements for the tests to be meaningful. The
results are presented in Table~\ref{tab:clasCompatibilityIR}.
The statistical tests indicate the following results:

\begin{itemize}
\item The spectral gradient distributions of the various resonant
  classes are statistically indistinguishable.  To increase
  the size of the sample, all the resonant objects were divided only
  between Plutinos (3:2) and ``other resonant objects''.
\item The spectral gradient distributions of Plutinos and other resonant
  objects, SDOs, hot CDOs, and Centaurs are indistinguishable.
\item Additionally, the Plutinos and the hot CDOs had indistinguishable
  $J-H$ distributions. 
\item The spectral gradient distribution of DDOs is indistinguishable
  from that of SDOs and only marginally distinguishable from those of
  Plutinos and resonant objects, as well as hot CDOs.  
\item The spectral gradient distribution of cold CDOs is 
  incompatible with the ones of hot CDOs, Plutinos, SDOs, DDOs, and
  Centaurs, and is only marginally compatible with that of the
  resonant objects.
\item Additionally, the cold CDOs $J-H$ distribution is incompatible with
  that of the Plutinos and the hot CDOs.
\item The spectral gradient distribution of Jupiter Trojans clearly
  indicates its complete incompatibility with all other MBOSS
  populations. 
\item The spectral gradient distribution of SPCs  significantly
  differs from those of the Jupiter Trojans and cold CDOs. It is only
  marginally different from those of the hot CDOs, Plutinos, and other resonant
  objects as well as Centaurs, and it cannot be distinguished from that
  of SDOs and DDOs.
\item The $J-H$ colours of the Centaurs are  incompatible with those of the 
Plutinos and the hot CDOs.
\end{itemize}

In conclusion, Plutinos, other resonant objects, hot CDOs, SDOs, and DDOs may
have rather similar spectral gradients in the visible, while
cold CDOs and Jupiter Trojans form two separate groups of
spectral properties in the visible wavelength range. The peak in the
spectral gradient distribution of hot CDOs seems to be shifted
slightly to higher reddening, although for the moment this result is
not yet at a statistically significant level. Short-period comets may
agree with the first populations mentioned above. 

\subsubsection{Absolute magnitude}

\begin{figure*}
   \centering 
   a.\includegraphics[width=9cm]{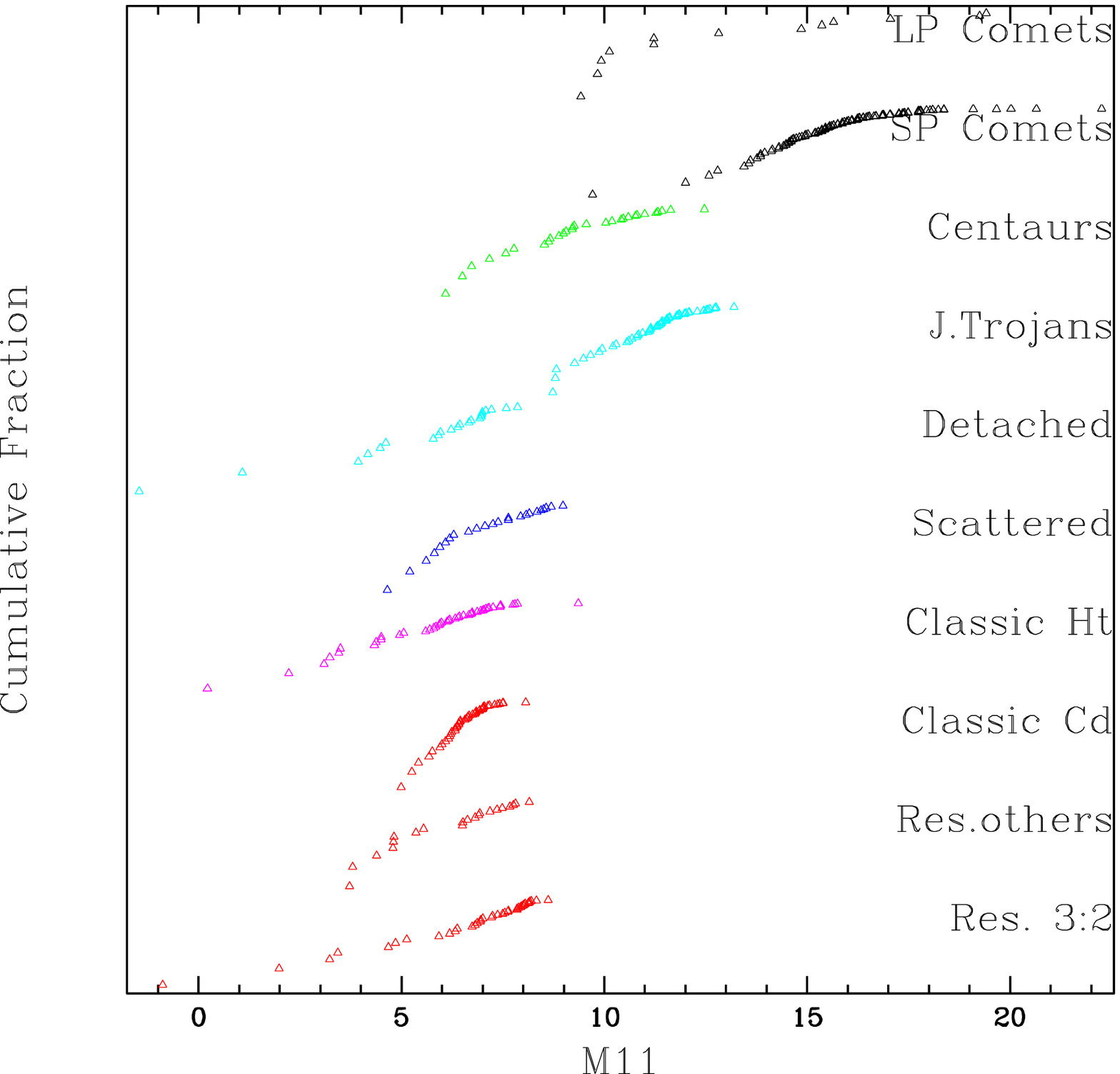}\hfill
   b.\includegraphics[width=9cm]{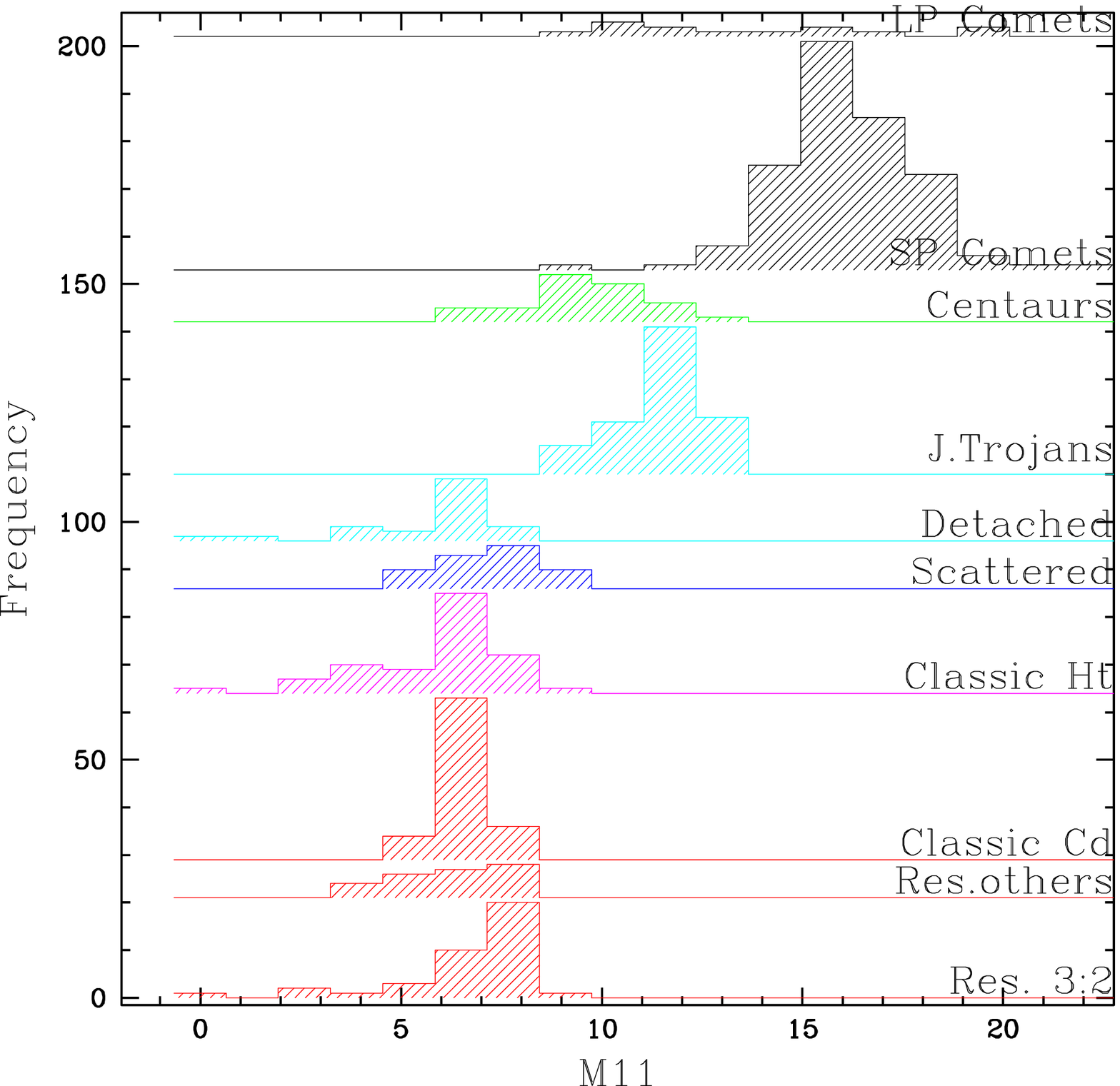}\\
   \caption{$R$ absolute mag
                            nitude $M(1,1)$ cumulative
     distributions (logarithmic, a), and histogram (b). }
   \label{fig:histM11}%
\end{figure*}

The absolute magnitude $M(1,1)$ of an object depends on the
reflectivity function across the surface, i.e. a first order
approximation on the object size, the geometric albedo, and the
photometric phase function. Given the relationship between $M(1,1)$,
size, albedo, and phase function, one can assume that the $M(1,1)$
distribution is dominated by the effects of the size, except for the
largest bodies, which often have high albedos \citep{Sta+08}. The
solar phase angle for KBOs is always relatively small, and the phase
function is moderately steep, so the effect on $M(1,1)$ hardly
exceeds 0.2-0.3mag. Hence, we assume that the $M(1,1)$ distribution of
KBOs reflects --at least to zeroth order-- the size distribution of
the objects, with the understatement that a detailed analysis requires
more accurate size estimates of the objects. Such a careful study can
be found in \cite{Sta+08}.

The $M(1,1)$ distributions are strongly
biased by cumulative selection effects: while there are few bright and
large objects in a given class, they are more easy to discover, and
also more easy to observe, so they are likely to be represented in
this data-set. A contrario, the discovery surveys are incomplete for
fainter objects, and faint objects are less likely to be picked by
observers performing colour measurements. Consequently, the $M(1,1)$
distributions are likely to be fairly complete at the bright end, but
have a complex non-completeness at the faint end. \cite{MHM04}
discussed and simulated many of these selection effects.

Fig. \ref{fig:histM11} shows the frequency histogram and cumulative
distribution function of $M(1,1)$ for the various MBOSS groups, now
covering all objects in the MBOSS database (i.e. not excluding those
with $M(1,1) > 5$~mag).

The selection bias against
small KBOs is clear, with no object fainter than $M(1,1)\sim 8.5$ in
the distant classes -- this corresponds to actual magnitudes fainter
than 23.5 for Plutinos, and even fainter for CDOs, i.e. about the
limiting magnitude for colour observations performed with fairly
short exposures on a 8~meter class telescope. On the other hand,
Centaurs and SPCs may be seen as a sample of smaller KBOs that is
scattered towards the inner planetary system. \cite{MHM04} found that
the shape of the SPC size distribution is incompatible
with that of the other TNOs.

The slope $k$ of the logarithmic cumulative $M(1,1)$ distribution
relates to the power law $q$ of the cumulative radius distribution
$N_c(a) \propto a^{-q}$ as $q = 5 k$. Measuring the $k$ slopes on the
straight parts of the distributions in Fig. \ref{fig:histM11} leads to
values $q \sim 0.8$ for the SPCs, the Centaurs, the faint SDOs, the
DDOs, the hot CDOs, and the resonant objects; the steep slope of the
cold CDOs and the bright SDOs corresponds to $q \sim 2$. Only the
steepest part of the SPCs distribution reaches $q\sim 3$. Analyses
based on controlled samples designed for size analyses have found much
steeper power-laws of from $q = 3.0$ \citep{TJL01} to $3.5$
\citep{KB04}, i.e. distributions much richer in small/faint
objects. This confirms that the sample used for colour measurements is
heavily biased against faint objects --the distributions may be
fairly complete only for the brightest SPCs.

Therefore, the size distributions in Fig. \ref{fig:histM11} should
only be used for internal comparison, with the hope that the selection
effects affected them in a similar way, and/or restricting
the analysis to the bright end of the distributions.

The observed Jupiter Trojans ($M(1,1)$ between 8 and 13~mag) are
clearly smaller than KBOs, which cover the $M(1,1)$ range from about 0
to 9~mag.  The bulk of that shift is caused by the proximity of the
Trojans, which enables objects fainter by $\sim 4$~mag to be observed.

It is noteworthy that cold CDOs display a rather peaked $M(1,1)$
distribution and do not contain objects with $M(1,1)$ brighter than
about 5~mag, while Plutinos, hot CDOs, and DDOs contain objects of
$M(1,1)$ from 2 to $-1$~mag. The rather steep slope in the cumulative
distribution function of $M(1,1)$ of the cold CDOs reflects the likely
difference in the size range of these KBOs. In particular, it differs
from that of hot CDOs, which should be exposed to very similar
selection effects, i.e. it is real. The brightest resonant objects
reach $M(1,1)$ of about 3mag, while the SDOs are more similar in
brightness to cold CDOs, despite having a much less pronounced peak
level on a slightly wider distribution plateau.

From the $M(1,1)$ distributions of MBOSSes, we conclude that the
absence of larger bodies among the cold CDOs and their presence among
other populations of the Kuiper Belt may suggest different formation
environments for both groups of objects that may have favored the
growth of larger bodies for Plutinos, SDOs, DDOs, and possibly hot
CDOs compared to cold CDOs. Jupiter Trojans are a population of
smaller size bodies, again arguing for a formation environment and
evolution scenario that were distinct from that of the other MBOSSes
(now including cold CDOs). Since it is very likely that the large
bodies in these groups may have survived impact events widely
unaffected over the lifetime of the solar system, one can assume that
they represent the original population of planetesimals. The Centaurs
reach the magnitude/size range where the vast majority of larger KBOs
are detected, but they do not include one of the rare big ones. This
might well be compatible with the scattering scenario for Centaurs but
needs quantitive confirmation by dynamical calculations.

Altogether, the $M(1,1)$ distributions give the impression that the
MBOSSes may have been formed in at least three (maybe four) different
environments, i.e. one for the Jupiter Trojans, one for the cold CDOs,
maybe one for the hot CDOs, and one for the rest of the bodies. Since
KBOs now reside and  for a very long time resided in the same distance
range, one may consider Plutinos, resonant objects, hot CDOs, and SDOs
as immigrant population of the Kuiper Belt compared to the cold CDOs
that may represent ``aborigines KBOs''. Interestingly, the DDOs also 
appear to be `sizewise' in accord with the Kuiper Belt ``immigrants'', 
although they are found at a much more distant location.

\subsection{Dependences on dynamical parameters}

\begin{figure}
   \centering 
   \includegraphics[width=9cm]{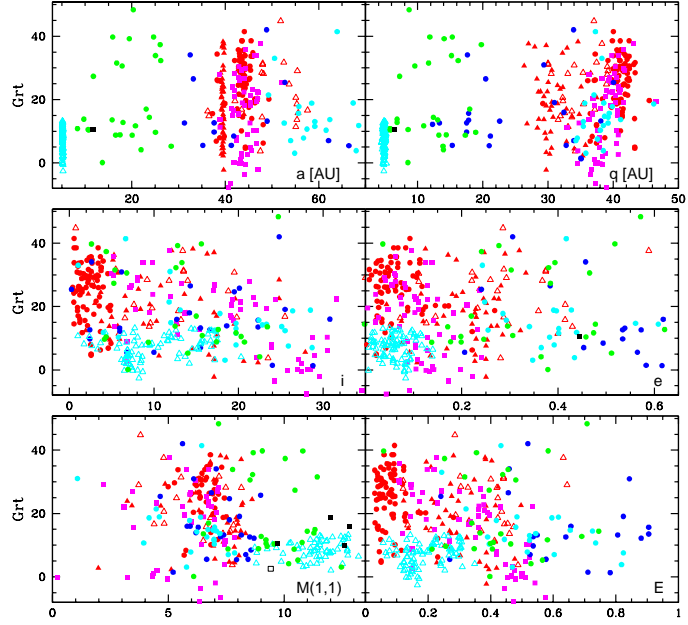}
   \caption{The spectral gradient of the objects [\%/100~nm] as a
     function of the main orbital elements, the orbital
     excitation $\cal E$, and the absolute $R$ magnitude
     $M(1,1)$. Similar plots are available on-line for the various
     colours.}
   \label{fig:orbitGrt}%
\end{figure}

The colour distribution (say, the $B-V$ distribution) of a class of
objects (say, the Centaurs) is considered as a function of another
parameter of the objects (say, the perihelion distance $q$) to search
for correlations as possible indicators of any physical
dependences. The traditional Pearson correlation factor is not robust
for non-linear dependencies, so a simpler and more robust estimator
was used: the sample is divided into two sub-samples, using the median
value of the independent variable (in the example, the median $q$ of
the Centaurs, i.e. 16.22~AU). The colour distribution of the two
sub-samples are then compared through $t$-test and $f$-test and the
Pearson correlation parameter is also estimated. Visually, this can be
done by comparing the left-hand side with the right-hand side of all
the sub-panels of Figure~\ref{fig:orbitGrt}.

\paragraph{For cold CDOs:}

The variances in the colours $B-V$, $B-R$ and $B-I$ are different for
the inner (smaller variance) and the outer (larger variance)
population among the cold CDOs (considering both their semi-major axis
and perihelion distance). These differences have a good level of
significance, as listed in Tab.~\ref{tab:coldCDOaq}. The effect is
only marginally observed in $V-R$ and $V-I$. The average colours of
the sub-samples are indistinguishable: only the breaths of the
distributions are different

Since similar results are found neither in other colours nor the
spectral gradients in the visible, the effect might be related to a
wavelength dependent phenomenon that is mostly restricted to the B
filter, for instance a stronger unknown continuum absorber in the
outer cold CDOs.

\begin{table}
\caption{Averge colours and dispersions for the cold CDOs, comparing
  those with smaller $a$ (or $q$) and those with large values. }
\label{tab:coldCDOaq}
\begin{center}
\begin{tabular}{crcrccc}
\hline
Colour  &  N  &  Aver./$\sigma$& N& Aver./$\sigma$&  $t$-Prob &$f$-Prob\\
  & \multicolumn{2}{c}{$a < 43.95$} & \multicolumn{2}{c}{$a \le 43.95$}\\
\hline
$B-V$&16 & 0.99$\pm$0.16 & 18& 1.03$\pm$0.11 & 0.148& 0.148\\
$B-R$&16 & 1.59$\pm$0.26 & 17& 1.69$\pm$0.09 & 0.151& 0.000\\
$B-I$& 9 & 2.18$\pm$0.37 & 14& 2.27$\pm$0.15 & 0.489& 0.003\\
$V-R$&21 & 0.63$\pm$0.10 & 21& 0.65$\pm$0.06 & 0.310& 0.034\\
$V-I$&17 & 1.20$\pm$0.16 & 20& 1.24$\pm$0.14 & 0.395& 0.538\\
\hline
  & \multicolumn{2}{c}{$q < 40.91$} & \multicolumn{2}{c}{$q \le 40.91$}\\
\hline
$B-V$&17 & 0.95$\pm$0.17 & 17& 1.07$\pm$0.06 & 0.013& 0.000\\
$B-R$&16 & 1.56$\pm$0.25 & 17& 1.72$\pm$0.07 & 0.024& 0.000\\
$B-I$&13 & 2.15$\pm$0.30 & 10& 2.34$\pm$0.13 & 0.060& 0.017\\
$V-R$&19 & 0.61$\pm$0.08 & 23& 0.67$\pm$0.08 & 0.033& 0.802\\
$V-I$&18 & 1.20$\pm$0.18 & 19& 1.25$\pm$0.10 & 0.278& 0.010\\
\hline
\end{tabular}
\end{center}

Notes: The $a$ and $q$ cut-off are set at the median value; N is the
number of measurements; $t$-Prob and $f$-Prob are the probabilities
that the two sub-samples are randomly extracted from the same
distribution, evaluated with the $t-$ and $f-$tests. The other colours
show insignificant differences.
\end{table}

\paragraph{For hot CDOs:}

Hot CDOs with a perihelion closer to the Sun are significantly bluer
than the ones that do not get close, see Tab.~\ref{tab:hotCDOq}, 
confirming earlier suggestions by \citet{HBO+04} based on a smaller
object sample. This   
result is visible in the $B-V$, $B-R$, and
$B-I$ colours (e.g. $B-V = 0.80 \pm 0.12$ for $q<39.4$~AU, and $B-V =
1.02\pm0.13$~mag for the others). This result is based on fairly small
samples, which decreases the strength of its significance.

\begin{table}
\caption{Average colours and dispersions for the hot CDO, comparing
  those with smaller  $q$ and those with large values. }
\label{tab:hotCDOq}
\begin{center}
\begin{tabular}{crcrccc}
\hline
Colour  &  N  &  Aver./$\sigma$& N& Aver./$\sigma$&  $t$-Prob &$f$-Prob\\
  & \multicolumn{2}{c}{$q < 39.39$} & \multicolumn{2}{c}{$q \le 39.39$}\\
\hline
$B-V$&10 & 0.80$\pm$0.12 & 12& 1.02$\pm$0.13 & 0.000& 0.795\\
$B-R$&10 & 1.28$\pm$0.21 & 12& 1.64$\pm$0.18 & 0.000& 0.666\\
$B-I$& 8 & 1.71$\pm$0.28 &  9& 2.37$\pm$0.27 & 0.001& 0.838\\
$V-R$&14 & 0.47$\pm$0.14 & 15& 0.57$\pm$0.15 & 0.071& 0.792\\
$V-I$&12 & 0.94$\pm$0.24 & 13& 1.15$\pm$0.24 & 0.037& 0.967\\
\hline
\end{tabular}
\end{center}

Notes: The  $q$ cut-off is set at the median value; N is the number
of measurements;  $t$-Prob and
$f$-Prob are the probabilities that the two sub-samples are randomly
extracted from the same distribution, evaluated with the $t-$ and
$f-$tests. The other colours show insignificant differences.
\end{table}

Hot CDOs with high inclinations or with large orbital excitation
parameters are bluer than the others, see Tab.~\ref{tab:hotCDOi}, a
finding seen for several colours ($B-V$, $B-R$, $V-R$, $V-I$) as well
as for the spectral gradient in the visible. The sample sizes are
small, but sufficient. This result was already known
\citep{TB02,Dor+08} and is generally expressed as an anti-correlation
between reddening of the objects and their orbital energy $\cal
E$. The usual interpretation scenario is colour changes due to
resurfacing by impact ejecta if during a collision between a hot CDO
and another KBO subsurface material with different colour properties
is ejected and deposited on the surface by impact excavations. Since
cold CDOs do not display such an anti-correlation despite of their
occupying the same distance range as the hot CDOs do, it may be more
an effect of orbital energy, i.e. collision energy rather than
collision frequency.

\begin{table}
\caption{Average colours and dispersions for the hot CDOs, comparing
  those with smaller  $i$ and those with large values. }
\label{tab:hotCDOi}
\begin{center}
\begin{tabular}{crcrccc}
\hline
Colour  &  N  &  Aver./$\sigma$& N& Aver./$\sigma$&  $t$-Prob &$f$-Prob\\
  & \multicolumn{2}{c}{$i < 17.0$} & \multicolumn{2}{c}{$i \le 17.0$}\\
\hline
$B-V$&14 & 0.99$\pm$0.16 &  8& 0.80$\pm$0.09 & 0.002& 0.096\\
$B-R$&14 & 1.60$\pm$0.21 &  8& 1.26$\pm$0.20 & 0.002& 0.940\\
$B-I$&11 & 2.14$\pm$0.33 &  6& 1.70$\pm$0.31 & 0.020& 0.978\\
$V-R$&15 & 0.61$\pm$0.08 & 14& 0.43$\pm$0.16 & 0.001& 0.012\\
$V-I$&13 & 1.20$\pm$0.15 & 12& 0.89$\pm$0.26 & 0.002& 0.064\\
Grad.&15 &24.59$\pm$7.99 & 13& 9.95$\pm$11.80& 0.001& 0.165\\
\hline
\end{tabular}
\end{center}

Notes: The  $q$ cut-off is set at the median value; N is the number
of measurements;  $t$-Prob and
$f$-Prob are the probabilities that the two sub-samples are randomly
extracted from the same distribution, evaluated with the $t-$ and
$f-$tests. The other colours show insignificant differences.
\end{table}

\paragraph{For others:} 
Other combinations of dynamical parameters and colours, spectral
gradient, and $M(1,1)$ data of different dynamical classes do not show
any statistically significant signal indicating possible correlations
and physical dependencies.

\subsection{Absolute magnitude}
\subsubsection{Absolute magnitude vs orbital parametres}
For the comparison between $M(1,1)$ and dynamical parameters, we
consider the “small” objects ($M(1,1) \ge 5$~mag) only and we find
that, for Centaurs, SDOs, and DDOs, when considered together, objects
with small semi-major axis $a$ and perihelion distance $q$ display a
significantly fainter absolute magnitude $M(1,1)$ than the distant
ones. While this is a statistically strong result, it disappears when
the three classes are considered separately. Thus, it can simply be
seen as the selection bias that makes the more distant and thus
fainter objects harder to measure than the closer ones.

\subsubsection{Absolute magnitude of the blue objects, vs red objects:}
We now compare the absolute magnitude $M(1,1)$ distribution of a class
of objects as a function of their colour, to explore whether the
redder objects have the same $M(1,1)$ distribution as the bluer
objects. Considering only the small objects ($M(1,1) \ge 5$~mag), the
blue hot CDOs and the blue resonant objects have fainter absolute
magnitudes than the redder objects of the same type (see
Tab.~\ref{tab:m11colour}).

\begin{table}
\caption{Average $M(1,1)$ and dispersions for the hot CDOs, comparing
  those with blue vs red colours. }
\label{tab:m11colour}
\begin{center}
\begin{tabular}{crrcrccc}
\hline Colour & Colour & N & Aver./$\sigma$& N& Aver./$\sigma$&
$t$-Prob\\ & cut-off & \multicolumn{2}{c}{Blue} &
\multicolumn{2}{c}{Red}\\ 
\hline \multicolumn{2}{l}{Hot TNOs}\\ \hline
$B-V$&0.92& 11& 6.92$\pm$0.67 & 11& 6.54$\pm$0.62 & 0.183\\ 
$B-R$&1.55& 11& 7.09$\pm$0.45 & 11& 6.37$\pm$0.64 & 0.007\\ 
$B-I$&1.92&  8& 7.13$\pm$0.52 &  9& 6.41$\pm$0.67 & 0.025\\ 
$V-R$&0.56& 14& 6.88$\pm$0.61 & 15& 6.34$\pm$0.62 & 0.026\\ 
$V-I$&1.10& 12& 6.87$\pm$0.61 & 13& 6.29$\pm$0.65 & 0.031\\ 
$R-I$&0.54& 12& 6.85$\pm$0.53 & 13& 6.30$\pm$0.72 & 0.039\\ 
\hline \multicolumn{2}{l}{Resonant objects}\\ 
\hline 
$B-V$& 0.88& 18& 7.61$\pm$ 0.81& 18& 6.79$\pm$ 0.78&0.004& \\
$B-R$& 1.40& 18& 7.61$\pm$ 0.81& 18& 6.79$\pm$ 0.78&0.004& \\
$B-I$& 1.89& 15& 7.40$\pm$ 0.84& 16& 6.79$\pm$ 0.81&0.052& \\
$V-R$& 0.56& 23& 7.48$\pm$ 0.72& 23& 7.04$\pm$ 0.88&0.073& \\
$V-I$& 1.06& 22& 7.42$\pm$ 0.72& 22& 7.08$\pm$ 0.91&0.170& \\
$R-I$& 0.51& 21& 7.47$\pm$ 0.78& 22& 7.01$\pm$ 0.83&0.065& \\
\hline
\end{tabular}
\end{center}

Notes: The objects with $M(1,1)<5$ were excluded from this test; the
colour cut-off is set at the median value for the considered colour; N
is the number of measurements; $t$-Prob are the probabilities that the
two sub-samples are randomly extracted from the same distibution,
evaluated with the $t-$test. The other colours show insignificant
differences.
\end{table}

\subsubsection{Others parameters}
 Other combinations of the absolute magnitude with other parameters do
 not show any statistically significant signal.


\section{Interpretation scenarios}\label{REF-4}

In the following discussion, we try to elaborate and and understand the
results from the statistical analysis described in Section 3 in the
framework of two rather different scenarios, i.e. surface colouration
by on-going external processes and surface colours as a result of
the primordial constitution of the bodies.

\subsection{Surface colouration by external processes}

Colour diversity of MBOSSes was originally attributed to the action of
external processes on their surfaces. These process included
resurfacing by impacts, which were expected to produce the
neutral-bluish colours of fresh ice excavated and/or spread over the
surface, and an irradiation process, producing red colours via
radiation aging \citep[eg][for a model]{LJ96a}. When restricting our
MBOSS database samples to objects with $M(1,1) \ge 5$~mag, as done
during most of our analysis, a third process, i.e. resurfacing due to
atmospheric deposits of material from intrinsic activity, was
considered less important for the subsequent discussion, since it
should only affect large-sized bodies, the smaller ones losing the
material into space.

A critical argument for significant changes in the surface
reflectivity due to impacts comes from a lightcurve analysis of
MBOSSes: asymmetric or partial impact resurfacing of the body surfaces
should cause colour variegations over the rotation of the bodies that
are, however, usually not seen \citep[][S.~Sonnett, priv. comm.,
  J.~L.~Ortiz, priv. comm.]{Sek+02}.  Nevertheless, other authors
report colour changes with rotation phase, e.g. for Haumea
\citep{LJP08}. Thus, strong variegation is certainly not widespread.

In the colouration scenario based on external processes, DDOs are
expected to represent the end state of the radiation reddening of the
surface, since one can assume that, for these objects, impact
resurfacing does not occurs at a significant level, simply because the
spatial number density of DDOs seems very low and collisions should
therefore happen extremely rarely over the lifetime of the solar
system. From our statistical analysis, we found that DDOs are not the
reddest objects among MBOSSes. Cold CDOs instead show by far the
reddest surface colours. To explain the smaller reddening of DDOs in
this colouring scenario, one has to assume that cold CDOs do not
represent the end state of radiation reddening (despite showing the
reddest colours), but high radiation doses, similar to those to which
DDOs are exposed, decrease the spectral gradients again. Laboratory
experiments \citep{deB+08} have found materials that display such 
behaviour, although the end state of these material has rather
neutral spectral gradients in the visible wavelength range. However,
it remains unclear whether these materials exist in DDOs and, if not,
whether the laboratory results apply more generally to more or even
all possible surface compounds considered to exist in DDOs. It is also
noteworthy to mention the coincidence of the peaks in the spectral
gradient distributions of DDOs as well as SDOs, Plutinos,
resonant objects, and Centaurs; this may be by chance despite
the different importance of collisional resurfacing for these groups.

The colour cascade, with peaks in the colour histograms decreasing in
reddening from cold CDOs to hot CDOs, then SDOs, Plutinos, resonant
objects, and parts of the Centaurs, must be attributable to the
increasing importance of impact resurfacing. However, Centaurs live in
a rather empty zone of the planetary system, hence impact resurfacing
must contribute less to their evolution. Here, the peak in the
spectral gradient diagram could be attributed to resurfacing from
intrinsic activity. The very red Centaurs instead represent the
inactive objects that are subject to only radiation reddening, though
not with extremely high doses (otherwise they could be expected to
show similar reddening to DDOs).  Among the six known Centaurs showing
cometary activity, currently only one has red colours.  A larger
sample of colours of cometary active Centaurs is therefore needed.


Jupiter Trojans differ in terms of their surface reddening from the
MBOSSes. It is remarkable that Jupiter Trojans do not show very `red'
surfaces as objects in all other MBOSS populations do, despite their
high exposure to radiation from the Sun (owing to their closer
distance by a factor of about 100).  Thus, we conclude that this
difference may be due to their having different material properties
from the other MBOSSes.

An interesting result is seen for the SPCs: the population's peak is
well within the range of those of SDOs, Plutinos, moderately red
Centaurs, as well as DDOs. Impact resurfacing should be negligible for
SPCs and instead resurfacing due to activity should be far more
important. On the other hand, owing to the fast, continuous mass-loss
on short time-scales (the lifetime for active SPCs is estimated to be
100\,000 years or less), the surface deposits should represent the
intrinsic colour of the cometary material. Radiation aging is
unimportant over the lifetime of SPC nuclei in the inner planetary
system since cometary activity will result in fast and continuous
resurfacing of the body. Hence, SPCs give the impression that their
nuclei may consist --on a global scale-- of material with rather
homogeneous colour properties. As SPCs to a large extent originate
from the Kuiper Belt \citep{GOM+08,MLG08} and because of their colour
similarity to the immediate dynamical relatives therein (SDOs,
Plutinos) and the Centaurs from where SPCs are captured, it is
attractive to assume that the colours of these relatives should also
be globally uniform throughout the bodies.  \citet{Jew02} discusses
the missing very red SPCs, suggesting that the red surfaces are
rapidly buried as water ice sublimation starts at $r<6$~AU.

\subsection{Global colouration in the formation era}

An alternative interpretation of the global results for the colour and
$M(1,1)$ distribution is that the measured colours actually reflect
the global characteristics of the objects, that is throughout the
object and not only to its (aged) surface. This hypothesis is
supported by the interpretation of SPC colours and the similarities of
the peaks in the colour distribution of the MBOSS populations. We note
that \citet{BSF11} introduced a similar hypothesis for the colour
diversity among Kuiper Belt objects based on sublimation properties of
super-volatile ices. We describe below what can be seen as arguments
in support of such a `primordial' colouration scheme for the MBOSSes,
where `primordial' indicates that the colours of the body materials
originate from a time before the body was formed, i.e. the formation
period of the planetary system, and have remained largely unaltered
ever since.

An additional assumption needed is that the surface materials are not
---or at least not by a large amount--- affected by external processes
such as reddening due to cosmic radiation. The global character of the
material colours also implies that impact resurfacing would reproduce
the original pre-impact surface colours of the bodies and that
large-scale colour diversities on the surface should not
occur. Intrinsic activity in MBOSSes will expel gaseous species and
solids of which at least the solids may resemble the original colours,
while the gases, if not lost to interplanetary space, may condense on
the body as surface frost, in which case the surface colour may
change. However, our selection criterion to use only objects with
$M(1,1) > 5$~mag for the statistical analysis should significantly
reduce the possibility that MBOSSes that may display intrinsic
activity `contaminate' the study results.

In the context of this primordial colouration scenario, the
findings from our statistical analysis of the MBOSS colours allow us to
draw some interesting conclusions:

\begin{itemize}
\item At least four different colour populations may have existed in the
  protoplanetary disk, i.e. the cold CDOs, the hot CDOs, the
  SDO/DDO/Plutino populations, and the Jupiter Trojans. The four
  populations are characterized by different surface colours and
  spectral gradients of decreasing amplitude in the visible wavelength
  region, but may have very similar colours in the NIR.

\item The colour diversities among MBOSS populations with different
  surface colours call for formation environments at different and
  disjunct distances from the Sun. Migration transport may have
  shifted the original populations to their current locations in the
  planetary system, that is have placed cold and hot CDOs in the same
  distance range. This is very much in line with the Nice model ---see
  Section 7.1 of \citet{MLG08} for an overall description, or
  \citet{Lev+08} for more details, which gives an integrated
  overview of the dynamical history of the outer solar system. In the
  Nice model, the cold CDOs form a distinct population formed at larger
  distances from the Sun than the other classes, and in which the hot
  CDOs are then implanted in the later stages of
  Neptune's outward migration. The Nice model also specifically
  accounts for the separate nature of the Jupiter Trojans
  \citep{Mor+05} and is able to produce DDOs from a population at
  closer distances to the Sun than the cold CDOs. The Nice model lists
  an impressive list of successes in reproducing the dynamical structure
  of the current solar system; the observed similarity of the SDO, DDO,
  and resonant populations could help refining the overall picture
  painted by the model. However, DDOs and hot CDOs may represent a
  challenge for migration scenarios, in particular if one wants to
  include their colour similarities with other MBOSS
  populations.

\item Similar colour properties among MBOSS populations at different
  locations in the outer planetary system may indicate a common
  formation region of the bodies. If so, this could apply to both
  Plutinos and DDOs, objects nowadays found at rather different
  distances from the Sun.  Migration (for the Plutinos) and scattering
  (for the DDOs) might again have been involved in separating the
  bodies formed in a common source region.

\item The migration and scattering processes have kept a population of
  the same formation origin together and shifted them to confined
  destination regions with no obvious signs of intermixing from other
  sides. This conclusion suggests that the migration and scattering
  processes in the outer solar system might not have been very violent
  since they have retained the colour characteristics as population
  entities.

\item The colour distributions for the populations with `temporary'
  dynamical object association, i.e. SDOs, Centaurs, and SPCs, should
  reflect the efficiency of the injection and ejection processes of
  bodies from the various feeding populations, i.e. Plutinos and both
  hot and cold CDOs.

\end{itemize}

There are supportive and also critical arguments from the statistical
analysis for the conclusions above. In particular, the similarities
and differences in the $M(1,1)$ distributions provide compatible
pieces of evidence: the disjunct $M(1,1)$ distributions of the Jupiter
Trojans and both the hot and cold CDOs \citep[the latter having been
  noted in various papers and summarized in][]{Dor+08} may call for
different formation environments at different solar distances, as also
suggested by \citet{Mor+08}. On the other side, the existence of large
objects with $M(1,1) < 3$~mag among Plutinos and DDOs constitutes an
independent supportive argument for a formation region that would be
common among these populations and distinct from the cold CDOs, as
noted by \citet{LS01} and developed by \citet{Gom03}. Despite the
possibly different formation environments of Plutinos and DDOs on the
one hand and hot CDOs on the other as suggested by their colour
differences, the presence of large objects ($M(1,1) < 3$~mag) in all
three populations suggests that the planetesimal formation and
destruction processes were similarly efficient. On the other side, no
obvious explanation ---except an ad-hoc assumption as an outcome of
the formation--- can be offered for the anti-correlation between
surface colours and excitation parameter as well as the inclinations
of the hot CDO population, since the migration process should be
`colour-blind'. While the existence of very red objects among hot
CDOs, DDOs, and Plutinos could be attributed to the contamination of
cold CDOs, the presence of bodies with blue, neutral surfaces in all
populations except for cold CDOs and DDOs may indicate that there are
colour contaminants either from the primordial environment or due
to dynamical injection during a later phase.

\section{A summarizing discussion}\label{REF-5}


The MBOSS database first published in 2003 has since been updated
using subsequently published magnitude and colour measurements of
TNOs, Centaurs, Jupiter Trojans, and both short- and long-period
comets. A qualifying criterion has been applied to the data in the
database in order to distinguish noisy measurement sets from more
consistent ones. we have performed a new statistical analysis of the
qualified colours, spectral gradient, and absolute $R$ magnitudes
$M(1,1)$ data in the database  with the aim of identifying
group properties among the various dynamical MBOSS populations,
i.e. for the hot and cold classical disk objects (CDOs), the scattered
disk objects (SDOs), the Plutinos, and resonant objects, as well as 
related objects outside the Kuiper belt such as detached disk objects
(DDOs), Centaurs, and SPCs, using a
classification based on the SSBN08 system \cite{GMV08}. Furthermore,
Jupiter Trojans are included in the analysis since, from all that is
known, these objects are unrelated to the MBOSSes, although
represent an abundant population at the border of this region in
the solar system. To avoid or at least to reduce the impact on
the surface colours due to intrinsic activity, the statistical sample
was restricted to objects with normalized (but not yet phase
corrected) brightness $M(1,1) \ge 5$~mag.

Our statistical analysis has uncovered indications of
indistinguishable and incompatible colour distributions and both the
differences and similarities of the $M(1,1)$ parameters among the
MBOSS populations. These results are consistent with a scheme of four
different colour (and $M(1,1)$) groups in the MBOSS regime, i.e. (1)
cold CDOs, (2) hot CDOs, (3) Plutinos, resonant objects, SDOs, DDOs,
Centaurs, and SPCs, and (4) Jupiter Trojans. Additional correlations
between photometric and dynamical parameters of the various
populations were searched for, but only a single well-established one
was found in the data of the hot CDOs, for which colour
anti-correlates with inclination and the collision energy parameter.

The statistical results have been discussed in the context of two basic
interpretation scenarios, i.e. by explaining the surface colours of the
MBOSSes  as being (1) due to the interplay of resurfacing of impact
ejecta (blue colour shift) and radiation reddening (red colour shift)
and (2) of primordial nature imposed by the material
properties of the formation environment. The two scenarios have the
principle capability to explain our findings based on the analysis of the
MBOSS database. However, they can also be challenged, each of them in a
different way. The resurfacing scenario has to work out whether and
under which assumption it is able to reproduce the actually observed
colour, spectral gradient, and $M(1,1)$ distributions of the MBOSSes
and in particular that it can explain the relatively moderate
reddening of the DDOs in this picture. In this respect, further lab
measurements of the radiation reddening of reference materials under
high dose levels could play an important role.

The primordial surface-colour scenario provides a good explanation of
the apparent uniformity of surface colours among SPCs despite the
cometary activity that should be able to excavate material layers of
different colours if they exist throughout the nuclei. The primordial
surface-colour scenario works with ad hoc assumptions about the
colour, spectral gradient, and $M(1,1)$ parameter distributions that
require justification. Brown et al. (2011) indicated some first
arguments to this respect. However, beyond that, the scenario
interpretation relies on the object migration for the transport of the
MBOSSes from their formation environment to the various regions where
they are found today. Object migration in the early planetary system
has already been introduced to explain the existence of the hot CDOs
in the immediate neighborhood of the cold CDOs in the Kuiper Belt.
More comprehensive models have been introduced (see \citet{Mor+08} and
\citet{GOM+08} for reviews). These dynamical models are ---at the
moment--- mostly `colour-blind' and `size-blind' apart from an
a-priori postulation of colour differences for hot and cold CDOs. Our
analysis of the MBOSS colours and $M(1,1)$ size parameters would
require at least three, possibly four, different source regions in the
primordial disk. It is not very plausible that the MBOSSes remained in
these original source regions, but it is instead very likely that they
were redistributed after formation to the locations in and out of the
Kuiper belt where we find them today. A comprehensive model scenario
is not yet available that is able to consistently explain the formation
environment, the physical properties and the dynamical history of the
MBOSS population. Nonetheless, the dynamics, colour,
and size distributions together (possibly also albedos) of the
measured populations are to be seen as important and constraining
benchmarks for these model scenarios ---not only qualitatively, but
also quantitatively.

In parallel, considerable progress has been achieved in the taxonomic
classification of the objects. This process aims to
sort the objects among categories defined by similar colour
properties \citep[independently of the physico-dynamical classes
  ---see][]{Bar+05b}, that is an approach orthogonal to the one
presented here, which aims to compare the photometric properties of
physico-dynamical classes. The multi-pronged approaches of dynamical
simulations and modelling, taxonomic classification, and photometric
analysis is moving towards an integrated understanding of the MBOSS
populations.

\begin{acknowledgements}
H. Boehnhardt wishes to thank the European Southern Observatory for
supporting this work during his research stay in September and
November 2011 at ESO Headquarters in Garching/Germany. We are grateful
for the very valuable comments and input provided by the anonymous
referee.  This research has made use of NASA's Astrophysics Data
System.

\end{acknowledgements}

\linespread{1.}

\bibliographystyle{aa} 
\bibliography{mnemonic,MBOSS2,allColors}

\Online
\begin{appendix} 
\section{Online material}

The following tables list the colours of the objects in the database
at the time this analysis was performed (Table A.1), as well as the
corresponding references (Table A.2). These tables do not appear in
the journal; are available electronically from the CDS via anonymous
ftp to cdsarc.u-strasbg.fr (130.79.128.5) or via
\href{http://cdsweb.u-strasbg.fr/cgi-bin/qcat?J/A+A/}{\tt
  http://cdsweb.u-strasbg.fr/cgi-bin/qcat?J/A+A/}.  The up-to-date
version of the database is available from
\href{http://www.eso.org/~ohainaut/MBOSS}{\tt
  http://www.eso.org/\~{}ohainaut/MBOSS}.

\addtocounter{table}{1}
\longtabL{1}{
\begin{landscape}
\begin{longtable}{lcccccccccc} 
\caption{Object class and colour data   \label{tabOnline:allColours}}\\
\hline\hline 
Object & Class$^{(1)}$&Epochs$^{(2)}$& M11 $\pm\sigma$& Grt$^{(3)}$ $\pm\sigma$& B-V $\pm\sigma$& V-R $\pm\sigma$& R-I $\pm\sigma$& V-J $\pm\sigma$& J-H $\pm\sigma$& H-K $\pm\sigma$ \\ 
\hline 
\endfirsthead
\caption{continued.}\\
\hline\hline
Object & Class$^{(1)}$&Epochs$^{(2)}$& M11 $\pm\sigma$& Grt $\pm\sigma$& B-V $\pm\sigma$& V-R $\pm\sigma$& R-I $\pm\sigma$& V-J $\pm\sigma$& J-H $\pm\sigma$& H-K $\pm\sigma$ \\ 
\hline
\endhead
\hline
\endfoot
1P/Halley               & SP Comet  &  2 &13.558 $\pm$ 0.596 & 4.266 $\pm$ 2.015 & 0.720 $\pm$ 0.040 & 0.410 $\pm$ 0.030 & 0.390 $\pm$ 0.060 &---  &---  &--- \\
2P/Encke                & SP Comet  &  6 &15.207 $\pm$ 0.596 & 5.117 $\pm$ 2.574 & 0.780 $\pm$ 0.020 & 0.424 $\pm$ 0.053 & 0.408 $\pm$ 0.060 &---  &---  &--- \\
4P/Faye                 & SP Comet  &  2 &16.020 $\pm$ 0.009 & 7.951 $\pm$ 0.000 &---  & 0.450 $\pm$ 0.040 &---  &---  &---  &--- \\
6P/dArrest              & SP Comet  &  3 &16.239 $\pm$ 0.009 &13.264 $\pm$ 2.147 & 0.770 $\pm$ 0.040 & 0.563 $\pm$ 0.067 & 0.450 $\pm$ 0.040 &---  &---  &--- \\
7P/PW                   & SP Comet  &  2 &15.185 $\pm$ 0.009 & 4.264 $\pm$ 2.620 &---  & 0.400 $\pm$ 0.050 & 0.410 $\pm$ 0.060 &---  &---  &--- \\
8P/Tuttle               & SP Comet  &  2 &12.799 $\pm$ 0.009 &15.869 $\pm$ 2.326 & 0.890 $\pm$ 0.040 & 0.530 $\pm$ 0.040 & 0.530 $\pm$ 0.060 &---  &---  &--- \\
9P/Tempel 1              & SP Comet  & 14 &14.560 $\pm$ 0.460 & 9.959 $\pm$ 0.826 &---  & 0.468 $\pm$ 0.010 & 0.469 $\pm$ 0.013 &---  &---  &--- \\
10P/Tempel 2             & SP Comet  & 10 &14.287 $\pm$ 0.304 &13.740 $\pm$ 1.578 & 0.800 $\pm$ 0.020 & 0.521 $\pm$ 0.037 & 0.520 $\pm$ 0.030 &---  &---  &--- \\
\end{longtable}
\end{landscape}
}
\centering
%
\begin{table*}
\caption{References used for each object,  and number of
  epochs.}
\label{tabOnline:photRef}
\begin{tabular}{p{5cm}p{5cm}l}
\hline
Object & References & Nr.\\
\hline
\hline
\object{C/1983~H1 IRAS-Araki-Alcock} & \citet{Lam+05} & 1\\
\object{C/1983~J1 Sugano-Saigusa-Fujikawa} & \citet{Lam+05} & 1\\
\object{C/1984~K1 Shoemaker} & \citet{Lam+05} & 1\\
\object{C/1984~U1 Shoemaker} & \citet{Lam+05} & 1\\
\object{C/1986~P1 Wilson} & \citet{Lam+05} & 1\\
\object{C/1987~A1 Levy} & \citet{Lam+05} & 1\\
\object{C/1987~H1 Shoemaker} & \citet{Lam+05} & 1\\
\hline
\end{tabular} 
\end{table*}

\addtocounter{table}{-1}
\begin{table*}
\caption{Continued}
\begin{tabular}{p{5cm}p{5cm}l}
\hline
Object & References & Nr.\\
\hline
\hline
               \object{28P/Neujmin~1}
          &           \citet{JM88}
\\
          &         \citet{Del+01}
\\
          &          \citet{MHM04}
\\
          &         \citet{Lam+05}
\\
          &          \citet{CHM87}
 &  11\\
\object{29P/Schwassmann-Wachmann~1}
          &      \citet{Bauer+03b}
\\
          &            \citet{J09}
\\
\hline
\end{tabular} 
\end{table*}

\addtocounter{table}{-1}
\begin{table*}
\caption{Continued}
\begin{tabular}{p{5cm}p{5cm}l}
\hline
Object & References & Nr.\\
\hline
\hline
               \object{75P/Kohoutek}
          &         \citet{Lam+05}
 &   1\\
                    \object{76P/West-Kohoutek-Ikemura}
          &         \citet{Lam+05}
 &   1\\
               \object{77P/Longmore}
          &         \citet{Lam+05}
 &   1\\
               \object{78P/Gehrels 2}
          &         \citet{Lam+05}
 &   1\\
                    \object{79P/du Toit-Hartley}
          &         \citet{Lam+05}
 &   1\\
                  \object{81P/Wild~2}
          &         \citet{Lam+05}
 &   1\\
               \object{82P/Gehrels~3}
          &         \citet{Lam+05}
 &   1\\
\hline
\end{tabular} 
\end{table*}

\addtocounter{table}{-1}
\begin{table*}
\caption{Continued}
\begin{tabular}{p{5cm}p{5cm}l}
\hline
Object & References & Nr.\\
\hline
\hline
                    \object{140P/BS}
          &         \citet{Lam+05}
 &   1\\
             \object{141P/Machholz2}
          &         \citet{Lam+05}
 &   1\\
                    \object{143P/KM}
          &            \citet{J02}
\\
          &          \citet{JSF03}
\\
          &         \citet{Lam+05}
 &   5\\
               \object{144P/Kushida}
          &         \citet{Lam+05}
 &   1\\
                    \object{147P/Kushida-Muramatsu}
          &         \citet{Lam+05}
 &   1\\
\hline
\end{tabular} 
\end{table*}

\addtocounter{table}{-1}
\begin{table*}
\caption{Continued}
\begin{tabular}{p{5cm}p{5cm}l}
\hline
Object & References & Nr.\\
\hline
\hline
              \object{1997~RT$_{5}$}
          &           \citet{G+98}
\\
          &           \citet{B+02}
\\
          &         \citet{BEN+11}
 &   4\\
              \object{1997~RX$_{9}$}
          &           \citet{G+98}
 &   1\\
             \object{1997~SZ$_{10}$}
          &           \citet{TR00}
 &   1\\
            \object{1998~FS$_{144}$}
          &           \citet{TR03}
\\
          &         \citet{BEN+11}
\\
\hline
\end{tabular} 
\end{table*}

\addtocounter{table}{-1}
\begin{table*}
\caption{Continued}
\begin{tabular}{p{5cm}p{5cm}l}
\hline
Object & References & Nr.\\
\hline
\hline
            \object{1999~RT$_{214}$}
          &         \citet{Ben+09}
 &   1\\
            \object{1999~RX$_{214}$}
          &        \citet{Peix+04}
\\
          &         \citet{BEN+11}
 &   3\\
            \object{1999~RY$_{214}$}
          &        \citet{Peix+04}
 &   2\\
             \object{1999~TR$_{11}$}
          &           \citet{TR00}
\\
          &         \citet{BEN+11}
 &   2\\
            \object{1999~XY$_{143}$}
          &         \citet{BEN+11}
 &   1\\
\hline
\end{tabular} 
\end{table*}

\addtocounter{table}{-1}
\begin{table*}
\caption{Continued}
\begin{tabular}{p{5cm}p{5cm}l}
\hline
Object & References & Nr.\\
\hline
\hline
            \object{2001~QD$_{298}$}
          &        \citet{Dor+05a}
 &   1\\
            \object{2001~QF$_{298}$}
          &         \citet{DOR+07}
\\
          &         \citet{For+04}
\\
          &         \citet{ADS+06}
\\
          &        \citet{Dor+05a}
 &   8\\
            \object{2001~QX$_{322}$}
          &          \citet{TRC03}
\\
          &         \citet{BEN+11}
\\
\hline
\end{tabular} 
\end{table*}

\addtocounter{table}{-1}
\begin{table*}
\caption{Continued}
\begin{tabular}{p{5cm}p{5cm}l}
\hline
Object & References & Nr.\\
\hline
\hline
     \object{5130-Ilioneus}         
          &         \citet{For+07}
 &   1\\
     \object{5145-Pholus}           
          &            \citet{B92}
\\
          &         \citet{Mue+92}
\\
          &         \citet{Dav+93}
\\
          &           \citet{D+98}
\\
          &           \citet{RT99}
\\
          &           \citet{W+97}
\\
\hline
\end{tabular} 
\end{table*}

\addtocounter{table}{-1}
\begin{table*}
\caption{Continued}
\begin{tabular}{p{5cm}p{5cm}l}
\hline
Object & References & Nr.\\
\hline
\hline
    \object{15760-1992~QB$_{1}$}    
          &           \citet{JL01}
\\
          &           \citet{TR00}
\\
          &         \citet{HBO+01}
\\
          &         \citet{BEN+11}
\\
          &           \citet{R+97}
 &   5\\
    \object{15788-1993~SB}          
          &          \citet{GHL01}
\\
          &           \citet{JL01}
\\
\hline
\end{tabular} 
\end{table*}

\addtocounter{table}{-1}
\begin{table*}
\caption{Continued}
\begin{tabular}{p{5cm}p{5cm}l}
\hline
Object & References & Nr.\\
\hline
\hline
    \object{18137-2000~OU$_{30}$}   
          &         \citet{For+07}
 &   1\\
    \object{18268-Dardanos}         
          &         \citet{For+07}
 &   1\\
    \object{18493-Demoleon}         
          &         \citet{For+07}
 &   2\\
    \object{18940-2000~QV$_{49}$}   
          &         \citet{For+07}
 &   1\\
    \object{19255-1994~VK$_{8}$}    
          &           \citet{TR00}
\\
          &         \citet{Dor+01}
\\
          &         \citet{BEN+11}
 &   4\\
\hline
\end{tabular} 
\end{table*}

\addtocounter{table}{-1}
\begin{table*}
\caption{Continued}
\begin{tabular}{p{5cm}p{5cm}l}
\hline
Object & References & Nr.\\
\hline
\hline
    \object{24952-1997~QJ$_{4}$}    
          &          \citet{GHL01}
\\
          &           \citet{JL01}
\\
          &         \citet{ADS+01}
\\
          &           \citet{B+02}
 &   4\\
    \object{24978-1998~HJ$_{151}$}  
          &           \citet{TR03}
\\
          &         \citet{BEN+11}
 &   2\\
    \object{25347-1999~RQ$_{116}$}  
          &         \citet{For+07}
 &   1\\
\hline
\end{tabular} 
\end{table*}

\addtocounter{table}{-1}
\begin{table*}
\caption{Continued}
\begin{tabular}{p{5cm}p{5cm}l}
\hline
Object & References & Nr.\\
\hline
\hline
    \object{33128-1998~BU$_{48}$}   
          &         \citet{ADS+04}
\\
          &         \citet{Dor+02}
\\
          &         \citet{ADS+06}
\\
          &      \citet{Bauer+03b}
 &   9\\
    \object{33340-1998~VG$_{44}$}   
          &         \citet{HBO+01}
\\
          &         \citet{Dor+01}
\\
          &         \citet{McB+03}
\\
\hline
\end{tabular} 
\end{table*}

\addtocounter{table}{-1}
\begin{table*}
\caption{Continued}
\begin{tabular}{p{5cm}p{5cm}l}
\hline
Object & References & Nr.\\
\hline
\hline
    \object{47171-1999~TC$_{36}$}   
          &         \citet{HBO+01}
\\
          &         \citet{ADS+01}
\\
          &         \citet{McB+03}
\\
          &         \citet{ADS+04}
\\
          &         \citet{Dot+03}
\\
          &          \citet{TRC03}
\\
          &         \citet{DOR+07}
\\
\hline
\end{tabular} 
\end{table*}

\addtocounter{table}{-1}
\begin{table*}
\caption{Continued}
\begin{tabular}{p{5cm}p{5cm}l}
\hline
Object & References & Nr.\\
\hline
\hline
    \object{55637-2002~UX$_{25}$}   
          &          \citet{JPH07}
\\
          &         \citet{DOR+07}
\\
          &          \citet{DM+09}
\\
          &          \citet{RST07}
 &   8\\
    \object{55638-2002~VE$_{95}$}   
          &          \citet{RST07}
\\
          &         \citet{BAR+06}
 &   2\\
    \object{56968-2000~SA$_{92}$}   
          &         \citet{For+07}
 &   1\\
\hline
\end{tabular} 
\end{table*}

\addtocounter{table}{-1}
\begin{table*}
\caption{Continued}
\begin{tabular}{p{5cm}p{5cm}l}
\hline
Object & References & Nr.\\
\hline
\hline
    \object{79360-1997~CS$_{29}$}   
          &           \citet{RT99}
\\
          &           \citet{JL01}
\\
          &         \citet{DAV+00}
\\
          &         \citet{HBO+01}
\\
          &         \citet{ADS+06}
\\
          &         \citet{BEN+11}
\\
          &          \citet{JPH07}
\\
\hline
\end{tabular} 
\end{table*}

\addtocounter{table}{-1}
\begin{table*}
\caption{Continued}
\begin{tabular}{p{5cm}p{5cm}l}
\hline
Object & References & Nr.\\
\hline
\hline
    \object{91133-1998~HK$_{151}$}  
          &         \citet{HBO+01}
\\
          &         \citet{McB+03}
\\
          &         \citet{Dor+02}
\\
          &         \citet{BEN+11}
\\
          &         \citet{Dor+01}
 &   6\\
    \object{91205-1998~US$_{43}$}   
          &        \citet{Peix+04}
 &   2\\
    \object{91554-1999~RZ$_{215}$}  
          &           \citet{B+02}
 &   1\\
\hline
\end{tabular} 
\end{table*}

\addtocounter{table}{-1}
\begin{table*}
\caption{Continued}
\begin{tabular}{p{5cm}p{5cm}l}
\hline
Object & References & Nr.\\
\hline
\hline
            \object{136472-Makemake}
          &          \citet{RST07}
 &   1\\
     \object{137294-1999~RE$_{215}$}
          &           \citet{B+02}
\\
          &         \citet{BEN+11}
 &   3\\
     \object{137295-1999~RB$_{216}$}
          &           \citet{B+02}
 &   1\\
      \object{138537-2000~OK$_{67}$}
          &         \citet{ADS+01}
\\
          &         \citet{ADS+04}
\\
          &         \citet{Dor+02}
\\
\hline
\end{tabular}
\end{table*}

\end{appendix}
\end{document}